\documentclass[
    prl,
    twoside,
    twocolumn,
    10pt,
    floatfix,
    showpacs,
    citeautoscript,
    superscriptaddress,
]{revtex4-2}

\usepackage{amsmath}
\usepackage{amssymb}
\usepackage{booktabs}
\usepackage{comment}
\usepackage{glossaries}
\usepackage{graphicx}
\usepackage[version=4]{mhchem}
\usepackage{siunitx}
\usepackage{tabularx}
\usepackage[dvipsnames]{xcolor}
\usepackage[colorlinks=true]{hyperref}

\hypersetup{
    pdffitwindow=false,
    pdfstartview={FitH},
    pdfnewwindow=true,
    colorlinks=true,
    linkcolor=Black,
    citecolor=Black,
    filecolor=Black,
    urlcolor=Black,
    pdftitle={Accelerating Plasmonic Hydrogen Sensors for Inert Gas Environments by Transformer-Based Deep Learning},
    pdfauthor={V. Martvall, H. Klein Moberg, A. Theodoridis, D. Tomeček, P. Ekborg-Tanner, S. Nilsson, G. Volpe, P. Erhart, and C. Langhammer},
}

\global\let\oldnewlabel\newlabel
\gdef\newlabel#1#2{\newlabelxx{#1}#2}
\gdef\newlabelxx#1#2#3#4#5#6{\oldnewlabel{#1}{{#2}{#3}}}
\let\newlabel\oldnewlabel
\graphicspath{{figures/}}

\setacronymstyle{long-short}
\newacronym{doe}{DOE}{U.S. Department of Energy}
\newacronym{lemas}{LEMAS}{Long Short-term Transformer Ensemble Model for Accelerated Sensing}
\newacronym{lod}{LOD}{limit of detection}
\newacronym{loq}{LOQ}{limit of quantification}
\newacronym{lspr}{LSPR}{localized surface plasmon resonance}
\newacronym{lstr}{LSTR}{long short-term transformer}
\newacronym{ml}{ML}{machine learning}
\newacronym{mlp}{MLP}{multi-layer perceptron}
\newacronym{rmse}{RMSE}{root-mean-square error}
\newacronym{sota}{SotA}{state of the art}

\DeclareSIUnit\angstrom{\text {Å}}
\DeclareSIUnit\hydrogenpercent{vol. \% \ \ce{H2}}
\DeclareSIUnit\bar{bar}

\renewcommand{\epsilon}[0]{\varepsilon}

\newcommand{\addchalmers}{
    Department of Physics,
    Chalmers University of Technology,
    SE-41296, Göteborg, Sweden
}
\newcommand{\addgu}{
    Department of Physics, University of Gothenburg, 412 96 Göteborg, Sweden
}

\begin{document}

\title{
   Accelerating Plasmonic Hydrogen Sensors for \texorpdfstring{\\}{}
   Inert Gas Environments by Transformer-Based Deep Learning
}

\author{Viktor Martvall}
\author{Henrik Klein Moberg}
\author{Athanasios Theodoridis}
\author{David Tomeček}
\author{Pernilla Ekborg-Tanner}
\author{Sara Nilsson}
\affiliation{\addchalmers}
\author{Giovanni Volpe}
\affiliation{\addgu}
\author{Paul Erhart}
\email{erhart@chalmers.se}
\author{Christoph Langhammer}
\email{clangham@chalmers.se}
\affiliation{\addchalmers}


\begin{abstract}
The ability to rapidly detect hydrogen gas upon occurrence of a leak is critical for the safe large-scale implementation of hydrogen (energy) technologies.
However, to date, no technically viable sensor solution exists that meets the corresponding response time targets set by stakeholders at technically relevant conditions.
Here, we demonstrate how a tailored \gls{lemas} accelerates the response of a state-of-the-art optical plasmonic hydrogen sensor by up to a factor of 40 in an oxygen-free inert gas environment, by accurately predicting its response value to a hydrogen concentration change before it is physically reached by the sensor hardware.
Furthermore, it eliminates the pressure dependence of the response intrinsic to metal hydride-based sensors, while leveraging their ability to operate in oxygen-starved environments that are proposed to be used for inert gas encapsulation systems of hydrogen installations.
Moreover \gls{lemas} provides a measure for the uncertainty of the predictions that is pivotal for safety-critical sensor applications.
Our results thus advertise the use of deep learning for the acceleration of sensor response, also beyond the realm of plasmonic hydrogen detection.
\end{abstract}

\maketitle

\section{Introduction}

The ability to detect, quantify and distinguish chemical species accurately and rapidly is crucial for technologies requiring swift data capture to support well informed decision-making, automation and process-monitoring.
Such technologies span a wide range of applications, including environmental monitoring \cite{ZHANG201087}, biosensing for real-time disease diagnostics \cite{ribet2018real}, chemical process control \cite{RolRudHub20} and food quality evaluation \cite{POGHOSSIAN2019111272}.
They all have in common that they critically rely on the development of sensors that are not only precise, sensitive and selective but also respond rapidly to their target substance and are able to deliver an accurate quantitative measure of the concentration of that target.

A domain that is rapidly expanding and where sensing will play a pivotal role in facilitating safe large scale implementation is hydrogen-based technologies, including fuel cells for heavy transport, shipping and aviation, energy storage solutions and green steel production.
They all have in common that they promise substantial reductions of greenhouse gas emissions.
However, this prospect also generates new demands for active process monitoring and control, and introduces safety concerns owing to the high flammability of \ce{H2}-air mixtures.
All of these issues can be effectively addressed by the development of fast and accurate \ce{H2} sensors.

From a sensing environment perspective, two distinct settings exist, where ambient conditions characterized by an abundance of oxygen constitutes the most obvious one. 
The second setting, which is of significant technological relevance but much less discussed in the scientific literature to date, is so-called ``inert'' or ``oxygen starved'' environments. 
They are established to encapsulate/enclose large scale \ce{H2} installations, such as entire engine rooms on fuel-cell powered ships, or fuel pipes on \ce{H2}-powered airplanes, to avoid the formation of flammable air-\ce{H2} mixtures. 
The rapid detection of even the tiniest \ce{H2} leaks inside these inert gas encapsulation infrastructures is critical to provide enough time for the implementation of appropriate measures to eliminate, as well as to spatially localize, the leak by placing a large number of sensors at strategic locations inside the system.
Specifically, in such installations the system is continuously flushed by an inert gas, such as \ce{N2} or \ce{Ar}, to eliminate the presence of molecular oxygen. 
Importantly, we note that the inert gas used in such systems will be of low quality from a purity perspective with respect to species such as \ce{H2O}, \ce{CO} or \ce{SO_x}, for cost reasons. 
This combination of lack of \ce{O2} and presence of sizable amounts of ``poisoning'' molecules that bind strongly to many sensor surfaces poses a significant challenge because (i) established \ce{H2} sensors of the catalytic and thermal type require \ce{O2} to work and (ii) because the strong molecular bonds either block/poison surface sites required for \ce{H2} dissociation and/or detection, or facilitate surface reactions that consume hydrogen species and thus prevent them from being detected \cite{DarNugLan20}.

To steer the development of next-generation \ce{H2} sensors that meet the upcoming demands of widely implemented \ce{H2} technologies outlined above, agencies and stakeholders have defined performance targets.
The most well-know ones are defined by the \gls{doe}, which identify sensor speed at ambient conditions as one of the key unresolved metrics \cite{doe-targets}.
To this end, a small number of studies exist where \ce{H2} sensors with response times just below \qty{1}{\second} for a \qty{1}{\milli\bar} \ce{H2} pulse have been demonstrated experimentally \cite{NugDarCus19, luong2021sub, BanSch21}.
However, while indeed important breakthroughs, these demonstrations were made in an idealized pure \ce{H2}-vacuum environment that constitutes a severe simplification.
As the main reason for this simplification we identify the aforementioned challenge of ``poisoning'' molecular species in technologically relevant sensing environments due to their impact on the surface chemistry of a sensor.
Hence, even though these demonstrations of \ce{H2} detection with sub-second response in idealized vacuum/pure \ce{H2} conditions exist, it is clear that further advances in this field are necessary \cite{DarNugLan20}.
 
Traditionally, such advances are attempted by developing new sensing materials, by nanostructuring the sensing materials and/or signal transducers, and by the refinement or modification of fundamental physical sensing mechanisms \cite{DarNugKad19, NugDarCus19, AlEbSa2020, KABCUM201676, HASSAN2016435, BanSch21, LoMaGu2021}.
Interestingly, however, only very limited attention has been directed towards harnessing the potential of tailoring the treatment of output data of existing sensor platforms with the aim to improve sensor response, e.g., by machine learning techniques.
Accordingly, only few recent studies have leveraged the potential of machine learning to enhance the \emph{accuracy} or \emph{sensitivity} of different kinds of gas sensors \cite{Huang2023, KanMinCho2022}, leaving the potential to enhance \emph{sensor response times} still largely unexplored.

In response to and motivated by the high demand for faster sensors in general and \ce{H2} sensors for inert gas environments in particular, here, we develop an approach for accelerating \ce{H2}-sensing that combines optical nanoplasmonic sensors based on hydride-forming metal nanoparticles, such as Pd and its alloys with coinage metals \cite{AiBinSun, DarNugLan20, WadSyrLan14}, that enable operation in oxygen-starved environments, with deep learning.
We show that this combination enables predicting the \ce{H2} concentration in an inert gas (far) more quickly than the conventional approach, which is limited by the need to reach full thermodynamic equilibration of the sensor after a change in \ce{H2} concentration and hampered by sensor deactivation effects due to the presence of molecular contaminants.

To analyze the output data of plasmonic hydrogen sensors, which typically consists of time series of scattering or extinction spectra in the visible light spectral range \cite{WadSyrLan14, DarNugLan20}, the current \gls{sota} analysis widely applied in the field collapses each such measured spectrum to a single spectral descriptor, such as the spectral peak position, the full-width half maximum or the centroid position \cite{DahTeg06}.
In other words, the (potentially) vast amount of information contained in the full spectrum is lost in this analysis.

To harness this information with the aim to accelerate plasmonic \ce{H2} sensor response in inert gas environment, here, we introduce \acrshort{lemas}, short for \acrlong{lemas}, which improves the sensor speed by learning the relationship between the time dependence of the full spectrum and the \ce{H2} concentration, while simultaneously assessing uncertainty in the model predictions through model ensembles.
The \gls{lstr} architecture consists of a long and short-term memory and has been demonstrated to be well suited for modeling long time sequences \cite{Xu2021}.
We demonstrate that \gls{lemas} reduces the response time of a \ce{Pd70Au30} alloy plasmonic \ce{H2} sensor by up to 40 times when exposed to a distinct \ce{H2} pulse down to \qty{0.06}{\hydrogenpercent} in an inert gas environment at atmospheric pressure, in a scenario simulating a sudden large leak.
Furthermore, we illustrate the ability of \gls{lemas} to rapidly discern and quantify slow gradual changes in \ce{H2} concentration from mere noise in a simulated scenario of detecting a small leak in an enclosed inert gas environment.
This ability is critical for detecting \ce{H2} at as low concentrations as possible as quickly as possible, allowing sufficient time to apply safety measures, such as system shutdown, before a safety-critical \ce{H2} concentration is reached.
Finally, as an ensemble model \gls{lemas}
enables one to obtain uncertainty estimates, which is of fundamental importance for safety-critical applications, including but not limited to \ce{H2} sensing.

\begin{figure}
    \centering
    \includegraphics[width=\linewidth]{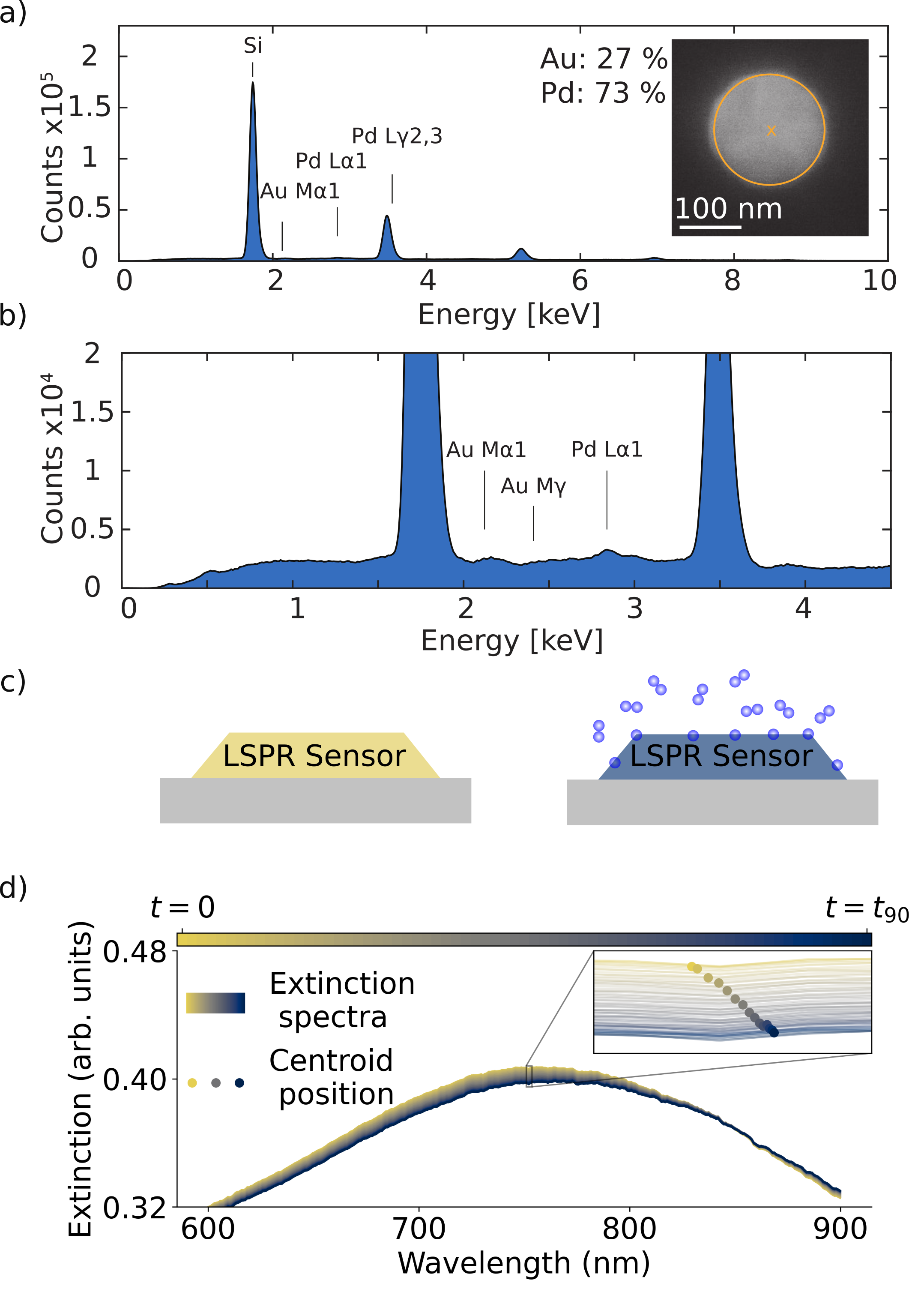}
    \caption{
        \textbf{\ce{Pd70Au30} alloy nanoparticle plasmonic sensor characterization and operating principle.} 
        (a) Energy-dispersive X-ray (EDX) spectrum collected from a single \ce{Pd70Au30} alloy nanodisk in the quasi-random array of such disks that constitutes the active sensor surface.
        %The particle is marked by the circle in the scanning electron microscopy image depicted in the inset.
        %The elementary composition of the alloy is obtained as an average of two circular scans together with one point scan in the center of the particle (marked with x in the inset).
        %We find reasonable agreement with the nominal composition of \ce{Pd70Au30}.
        (b) Zoom-in of the EDX-spectrum in (a) up to \qty{4.5}{\kilo\electronvolt} to focus on the characteristic Pd and Au peaks.
        %The peak at \qty{3.5}{\kilo\electronvolt} is due to pulse pile up, and is not included in the quantification fit used to derive the alloy composition.
        (c) Schematic illustration of the plasmonic \ce{H2} sensing principle, where the sorption of hydrogen into hydride-forming metal nanoparticles induces a change in their localized surface plasmon resonance frequency, which leads to a color change that is resolved in a spectroscopic measurement in the visible light spectral range.
        (d) Example of the spectral response of the \ce{Pd70Au30} alloy plasmonic sensor used in this work, resolved as a gradual shift in the extinction spectrum as hydrogen is absorbed the crystal lattice.
        Inset: Temporal evolution of the peak centroid position, one of the spectral descriptors that can be tracked to enable real time \ce{H2} detection.
    }
    \label{fig:sota_tem}
\end{figure}

\section{Results and discussion}

\paragraph{\ce{Pd70Au30} alloy plasmonic \ce{H2} sensors.}
As the plasmonic \ce{H2} sensor platform of choice, we selected the well-established \ce{Pd70Au30} alloy system, which we have investigated in detail earlier \cite{WadNugLid15, BanNugSch19, NugDarZhd18, NugDarCus19, EkbRahRos22}.
This material system is especially suited for inert gas sensing environments since its sensing mechanism, the interstitial sorption of hydrogen into the lattice of the metal host, does not require \ce{O2} to be present.
The Au alloyant serves the purpose of eliminating the intrinsic hysteresis characteristic for pure Pd by lowering the critical temperature of the system \cite{LeeNohFla07, LuoWanFla10, MamZhd20, RahLofFra21}.
At 30\% Au the best compromise between completely eliminating hysteresis, establishing linear optical response to \ce{H2} and maximizing optical contrast per unit sorbed \ce{H2} is reached.
Therefore, we nanofabricated quasi-random arrays of \ce{Pd70Au30} alloy nanodisks with a mean diameter of \qty{210}{\nano\meter} and \qty{25}{\nano\meter} height onto fused silica substrates using hole-mask colloidal lithography (\autoref{fig:sota_tem}a-b), following the procedures described in detail in our earlier work \cite{Nugroho2016} and in \autoref{sect:method-sample} in \autoref{sect:methods}.

The working principle of plasmonic \ce{H2} sensors is based on the \gls{lspr} phenomenon, characteristic for metal nanoparticles irradiated by visible light.
In an optical transmission, scattering or extinction spectrum, the \gls{lspr} manifests itself as a distinct peak with a maximum at a specific wavelength.
The spectral position of this peak maximum, as well as related peak descriptors such as width and intensity, exhibits a linear dependence on the \ce{H2} partial pressure surrounding the particles and on the amount of hydrogen species absorbed into interstitial lattice sites of the Pd or Pd alloy host (\autoref{fig:sota_tem}c-d) \cite{NugDarZhd18}.
Since the ab- and desorption of hydrogen into and from these interstitial lattice positions, respectively, occurs spontaneously and reversibly at ambient conditions, and also in oxygen-free environments, tracking of the spectral position (as well as other peak descriptors) of the \gls{lspr} peak as function of \ce{H2} partial pressure enables real time \ce{H2} detection (inset in \autoref{fig:sota_tem}d).
In this work, for what we refer to as the \gls{sota} analysis, we use the centroid position as a spectral descriptor which we relate to the \ce{H2} concentration by a calibration function (see \autoref{sect:method-sota} in \autoref{sect:methods} for details).

\begin{figure*}
\centering
\includegraphics[width=\linewidth]{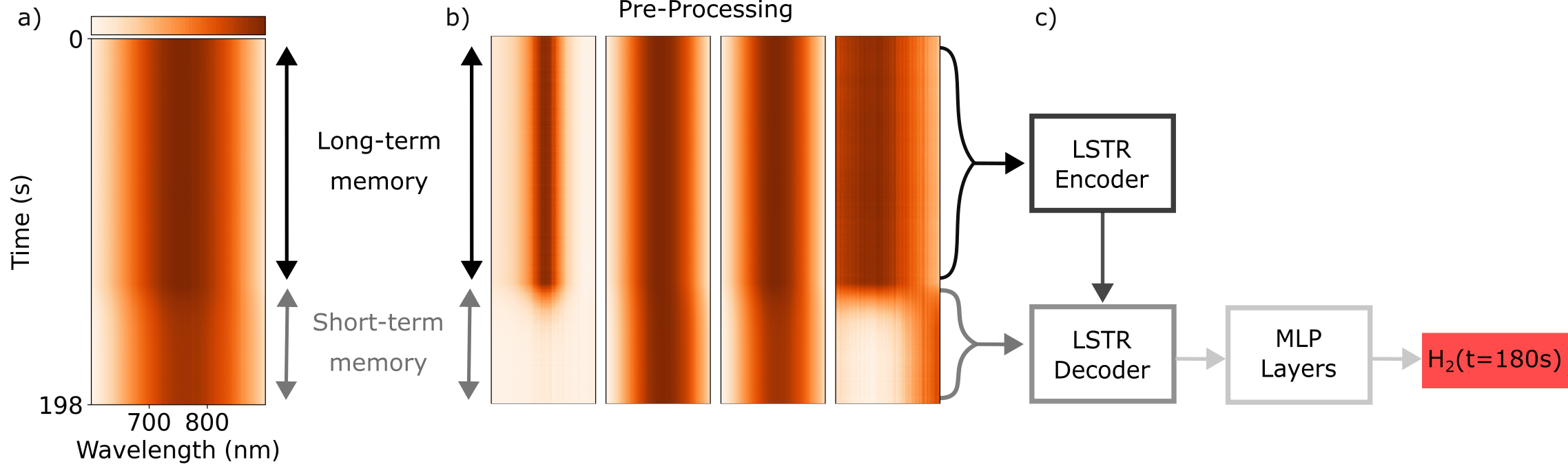}
\caption{
    \textbf{\acrlong{lemas}.} 
    Illustration of the deep learning model used in this work based on the \gls{lstr} architecture.
    (a) The input data to the model consists of a time sequence of the past evolution of the spectral response of the sensor.
    In this figure the time sequence consist of \num{600} time steps corresponding to \qty{198}{\second}.
    (b) The time sequence is first split into a long and short-term memory and pre-processed using four different methods, including wavelength dependent min-max normalization, standard normal variate standardization, global min-max normalization, and level scaling, before the concatenation of the pre-processed data is being fed to the \gls{lstr}.
    (c) The \gls{lstr} firstly compresses the long-term memory to a fixed length latent representation in the \gls{lstr} encoder.
    Secondly, the \gls{lstr} decoder extracts relevant temporal features in the short term memory while also querying the compressed long-term memory.
    The extracted temporal features are then passed through a stack of \gls{mlp} layers to obtain a prediction of the current \ce{H2} concentration.
}
\label{fig:lstr}
\end{figure*}

\paragraph{Deep learning model selection.}

We base our choice of a \gls{lstr} model for accelerating the plasmonic \ce{H2} sensor response on several key characteristics of the output data generated by this type of sensor (see \autoref{fig:lstr} and \autoref{sect:method-model} in \autoref{sect:methods} for details).
The first important characteristic to take into account is that the measured extinction spectra that constitute the raw sensor response over time, exhibit intrinsic noise (due to intensity fluctuations of the halogen light source and detection noise of the spectrometer used) that is comparable to the magnitude of changes induced in the spectra by small variations in \ce{H2} concentration.
Consequently, a crucial criterion for selecting the deep learning model is its ability to accurately model long temporal sequences, enabling the differentiation between relevant temporal trends in the extinction spectrum and the inherent noise.

The second critical aspect influencing the performance of the \gls{lstr} model, based on the characteristics of the sensor data, is the pre-processing of the measured extinction spectra.
Such pre-processing is needed due to drift in the sensor response over time (mainly due to long-term variations of light source intensity), as well as small variations in the extinction spectra obtained in different measurements using the same sensor, due to slightly different placement of the sensor in the measurement chamber for each independent experiment.
Here, we found that using several pre-processing methods is beneficial for the performance of the \gls{lstr} model.
Therefore, we used four different pre-processing techniques (see \autoref{sect:method-data-pre-proc} and \autoref{snote:data-pre-processing} for details) and concatenated them into a single array \cite{MISHRA2022116804}.
As a result, the input data for the deep learning models was a time series, where each element in the sequences consisted of the concatenation of the different pre-processing techniques (see \autoref{fig:lstr}a).

Another modeling choice that we make is to employ an \emph{ensemble} of \gls{lstr} models.
This choice is motivated by the safety-critical nature of the hydrogen sensor application and yields a more robust prediction, as well as a measure of uncertainty by aggregating the predictions of several \gls{lstr} models to compute the mean and the standard deviation (see \autoref{sect:method-ensembles} in \autoref{sect:methods} for details).
Combining these modeling choices, we arrive at \gls{lemas}, characterized by an ensemble of \gls{lstr} models that can both rapidly predict the \ce{H2} concentration and provide a measure of uncertainty from a time series of pre-processed spectra.

\paragraph{LEMAS model training and testing.}
Having introduced the architecture of the \gls{lemas} model, we discuss the training and testing data used for optimizing the sensor response in (i) a large and fast leak scenario and (ii) a slow gradual leak scenario.
These data were generated by measuring tailored time series of optical extinction spectra of the sensor localized in a custom-made measurement chamber with small volume to enable rapid gas exchange at atmospheric pressure to expose the sensor to varying  \ce{H2} concentrations in inert gas environment (see \autoref{sect:method-measurements} in \autoref{sect:methods} and \autoref{sfig:minireactor} for details).
Specifically, we used three different \ce{H2} profiles for generating the training data: (i) step-wise increase/decrease of \ce{H2} from \qty{0.00}{\hydrogenpercent} to 0.06--\qty{1.97}{\hydrogenpercent} in inert Ar environment (\autoref{sfig:step protocols}), (ii) linear increase/decrease of \ce{H2} from \qty{0.06}{\hydrogenpercent} to 0.09--\qty{1.97}{\hydrogenpercent} in inert Ar environment (\autoref{sfig:leak protocols}a) and (iii) exponential increase/decrease of \ce{H2} from \qty{0.06}{\hydrogenpercent} to 0.09--\qty{1.97}{\hydrogenpercent} in inert Ar environment (\autoref{sfig:leak protocols}b), see \autoref{snote:protocols} for details.

For the first case of a large simulated leak characterized by a rapid step-wise increase of \ce{H2} concentration in the sensor surroundings, we trained \gls{lemas} using two independent measurements of step-wise \ce{H2} concentration increase/decrease and subsequently tested the trained \gls{lemas} model on a third measurement not used for training.
For the second case of a simulated small slow leak, we trained \gls{lemas} on one measurement of linear \ce{H2} concentration increase/decreases and tested the performance of the trained model on one measurement with exponential \ce{H2} increase/decreases (see \autoref{sect:method-deep-learning-training} for details).

\begin{figure}
    \centering
    \includegraphics[width=0.8\linewidth]{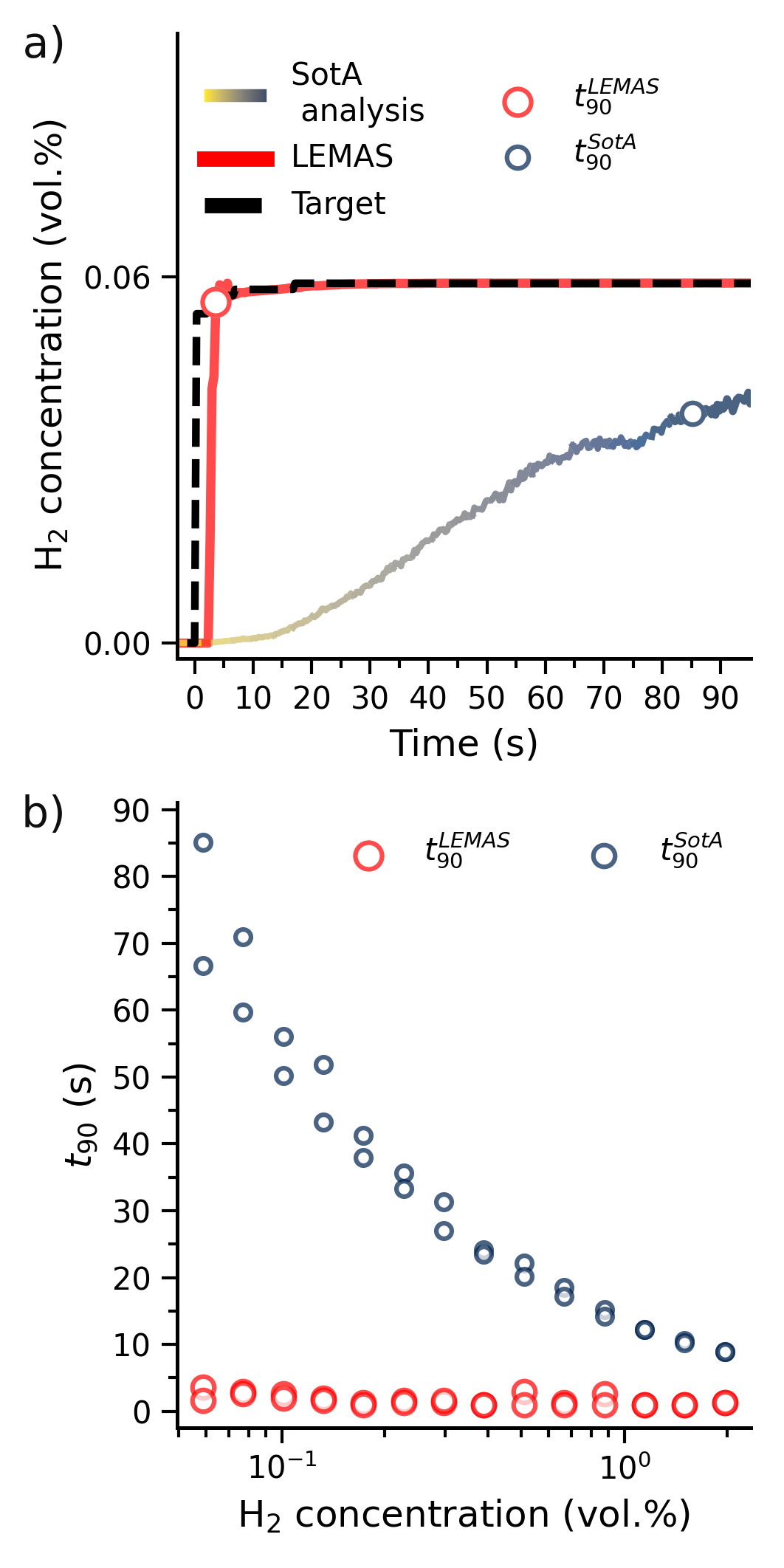}
    \caption{
        \textbf{Accelerating sensor response to a simulated large leak in inert gas environment.}
        (a) Comparison of the prediction of \gls{lemas} and the \acrshort{sota} centroid analysis for a pulse of \qty{0.06}{\hydrogenpercent} in inert Ar environment.
        By utilizing the full time-dependent spectrum of the measured sensor response \gls{lemas} is able to accurately predict the final value \ce{H2} concentration before the sensor physically reaches its new state in equilibrium with the new \ce{H2} concentration level.
        (b) Comparison of response times obtained by \gls{lemas} and the \acrshort{sota} centroid analysis as a function of \ce{H2} concentration in inert Ar environment.
        Note the significant acceleration by \gls{lemas}, in particular at the lowest \ce{H2} concentrations, and the elimination of the concentration dependence of the response.
    }
    \label{fig:sota_lemas}
\end{figure}

The models trained for optimizing the sensor response in a large and fast leak scenario used a total input sequence length corresponding to the past \qty{3}{\minute}, whereas the models trained for optimizing the sensor response in a slow gradual leak scenario used an input sequence length corresponding to the past \qty{22}{\minute} of the sensor history.
These choices were made based on an analysis of the change in centroid position in the training data and an estimation of the length of the time sequence needed to differentiate the slowest occurring process in the sensor output data from the noise in the measurement (see \autoref{sfig:timeseries_steps} and \autoref{sfig:timeseries_linear} for details).

\paragraph{Accelerating sensor response to a simulated large leak in inert Ar environment.} 
To assess the ability of \gls{lemas} to accelerate the response of a plasmonic \ce{H2} sensor, we first consider a scenario where a 0.06\% \ce{H2} pulse in inert Ar gas is applied to our device at \qty{30}{\celsius} (\autoref{fig:sota_lemas}a).
For this analysis, we define the response time $t_{90}$ as the first point in time where the sensor response has reached 90\% of its new steady state value.
Applying first the \gls{sota} analysis that tracks the centroid position, reveals that it takes on the order of \qty{85}{\second} to reach $t_{90}$.
Deploying the \gls{lemas} analysis on the same data, shows that it is able to predict the saturated \ce{H2} level after only  \qty{3.6}{\second} and thus long before the response of the system has saturated, leading to a more than 20-fold reduction of the response time.
This result is corroborated when comparing $t_{90}$ values obtained by \gls{sota} and \gls{lemas} analysis across a range of \ce{H2} pressure pulses from \qtyrange[range-phrase=~to~]{0.06}{1.97}{\hydrogenpercent} (\autoref{fig:sota_lemas}b and \autoref{sfig:SI_all-pulses_sota_ml}).

Remarkably, \gls{lemas} also achieves a response time that is practically independent of \ce{H2} concentration whereas the $t_{90}$ from the \gls{sota} analysis quickly increases with decreasing \ce{H2} concentration.
We attribute this behavior to the fact that \gls{lemas} only requires a certain number of data points to make its prediction, the availability of which is dictated by the read-out frequency of the spectrometer rather than the \ce{H2} pressure.
By contrast, the \gls{sota} analysis is limited by the intrinsic kinetics of the material platform, causing a strong dependence on the \ce{H2} pressure.
As a result, \gls{lemas} yields the largest boost in acceleration in the application critical range of lower \ce{H2} pressures and \gls{lemas}, overcoming one of the most important intrinsic limiting factors of hydride-based \ce{H2} sensors.

\begin{figure*}
    \centering
    \includegraphics[width=\linewidth]{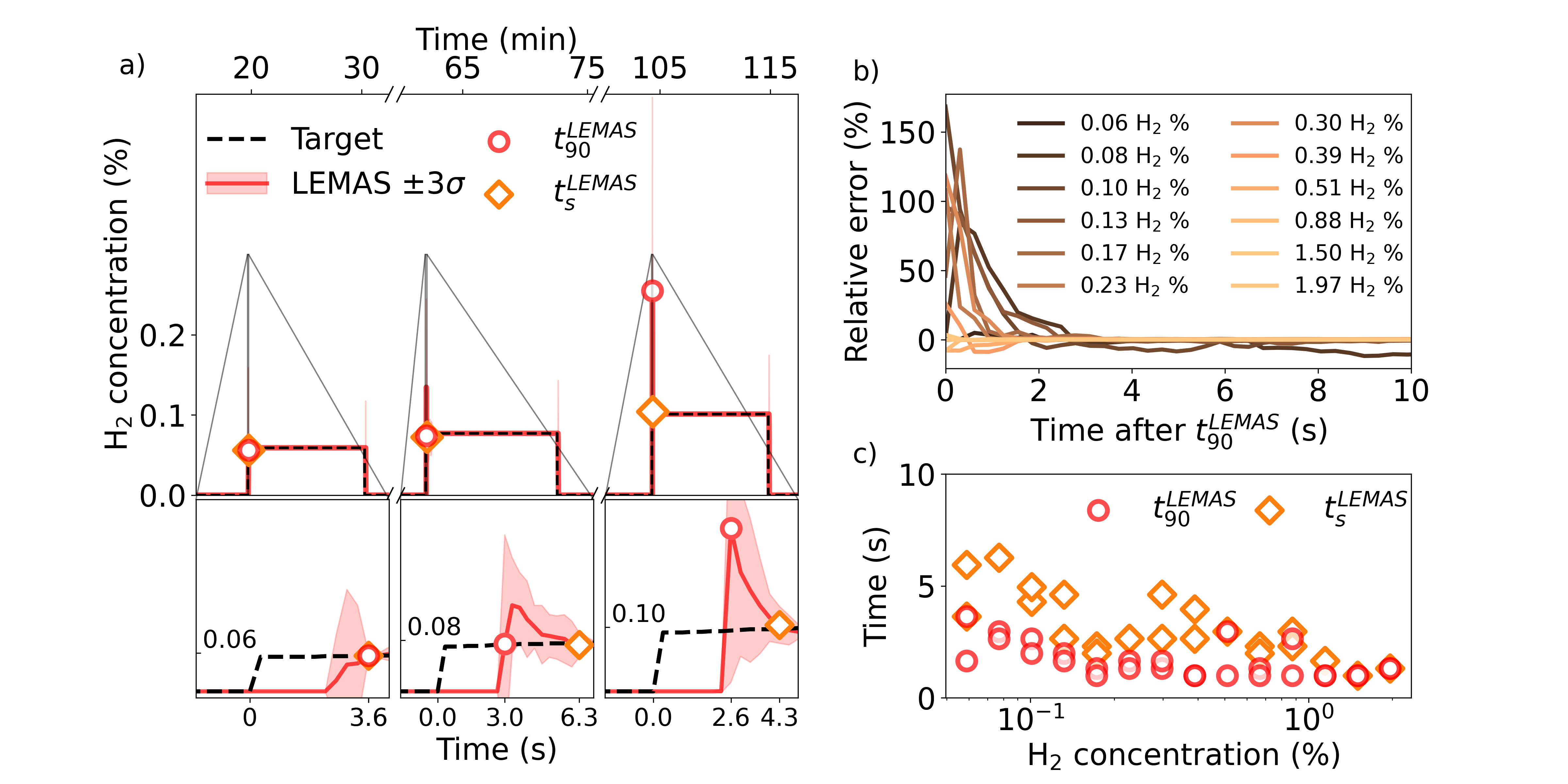}
    \caption{
        \textbf{\Gls{lemas} prediction accuracy to \ce{H2} pulses of different concentration.} 
        (a) The prediction and standard deviation from \gls{lemas} for three selected \ce{H2} concentration pulses in the test set.
        The lower panels display zoom-ins on the initial response to the pulse.
        (b) The relative error of the \gls{lemas} predictions for the entire range of \ce{H2} concentration pulses, starting at the response time, $t_{90}^\text{LEMAS}$, and forward.
        (c) \Gls{lemas} settling and response times as a function of the \ce{H2} concentration.
    }
    \label{fig:stepwise_accuracy}
\end{figure*}

Specifically, for the smallest concentrations considered in our experiment, at \qty{0.1}{\hydrogenpercent} and below, the response times range between \qtyrange[range-phrase=~--~]{1.6}{3.6}{\second} for the \gls{lemas} analysis compared to \qtyrange[range-phrase=~--~]{50}{85}{\second} for the \gls{sota} analysis.
This corresponds to an 21 -- 40 fold improvement compared to the \gls{sota}.
At the same time, we also note that even the accelerated response obtained by \gls{lemas} in the present inert gas conditions, is slower than the state-of the art in vacuum/\ce{H2} environment without acceleration \cite{NugDarCus19, luong2021sub, BanSch21}.
As the main reasons, we identify the following points:
(i) The traces of poisoning species such as \ce{H2O}, \ce{CO}, etc. present in the Ar inert gas used, significantly decelerate the sensor, as expected\cite{NugDarCus19} (see \autoref{sfig:gas_flow_1} and \autoref{sfig:gas_flow_2} for quantitative mass spectrometric analysis of the background molecular species present in the Ar inert gas used).
(ii) We have used relatively large nanoparticles, and it is known that reducing size increases sensor speed due to reduced hydrogen diffusion path lengths\cite{NugDarCus19}.
(iii) We have not applied any polymer coatings, which are known to accelerate sensor response, as well as protect them from the poisoning molecular species present in the inert gas \cite{NugDarCus19}.

Furthermore, we highlight that the amount of response time acceleration that \gls{lemas} can produce not only depends on the obvious intrinsic response speed of the active sensor material (in our case the \ce{PdAu} alloy nanoparticles) but also on the sampling rate of the sensor hardware, where a higher sampling rate enables a larger degree of acceleration.
For our experiments discussed so far, we have used a sampling frequency of \qty{3}{\hertz}, which was the the highest rate enabled by the used spectrometer.
Consequently, in this specific implementation, \gls{lemas} has only three data points available to identify a change in the \ce{H2} concentration in less than \qty{1}{\second}.
Crucially, the acceleration observed in the present case is thus not limited by \gls{lemas} but the underlying materials and read-out of the used light sampling device.

Having established the overall \gls{lemas} concept and demonstrated its ability to substantially accelerate sensor speed in inert sensing environments, in particular in the low concentration regime, it is interesting to evaluate the performance of \gls{lemas} in more detail.
To do so, we select three different \ce{H2} concentration pulses, i.e., pulses to 0.06, 0.08, and \qty{0.10}{\hydrogenpercent}, and plot the the sensor response predicted by \gls{lemas} as a function of time, with the standard deviation of the prediction at each time point indicated in the corresponding graphs (\autoref{fig:stepwise_accuracy}a and \autoref{sfig:SI_all-pulses} for all pulses).
We also define the sensor settling time, $t_{s}^\text{LEMAS}$, as the first time point where the predicted \ce{H2} concentration lies within $\pm$ \qty{10}{\percent} of the target \ce{H2} concentration and the relative standard deviation is smaller than \qty{10}{\percent}.
This metric complements the response time by also considering cases where \gls{lemas} either underestimates or overestimates the \ce{H2} concentration after $t_{90}$.

First, we note that at the onset of each pulse, there is a brief interval where \gls{lemas} predicts \qty{0}{\hydrogenpercent}, while the actual \ce{H2} concentration has already increased.
This behavior occurs since the change in the extinction spectrum induced by the presence of \ce{H2} is not yet distinguishable from the noise level in the measurement.
This initial phase is followed by an interval where a clear change in the extinction spectrum is detected but where both error (\autoref{fig:stepwise_accuracy}b) and uncertainty are still rather large (red shaded areas in \autoref{fig:stepwise_accuracy}a).
In the final phase, the \gls{lemas}-predicted \ce{H2} concentration settles at the correct value once the change in the extinction spectrum is sufficiently distinct, such that all models in the ensemble predict a similar \ce{H2} concentration, and the uncertainty becomes very small.

\begin{figure*}[tb]
    \centering
    \includegraphics[width=\linewidth]{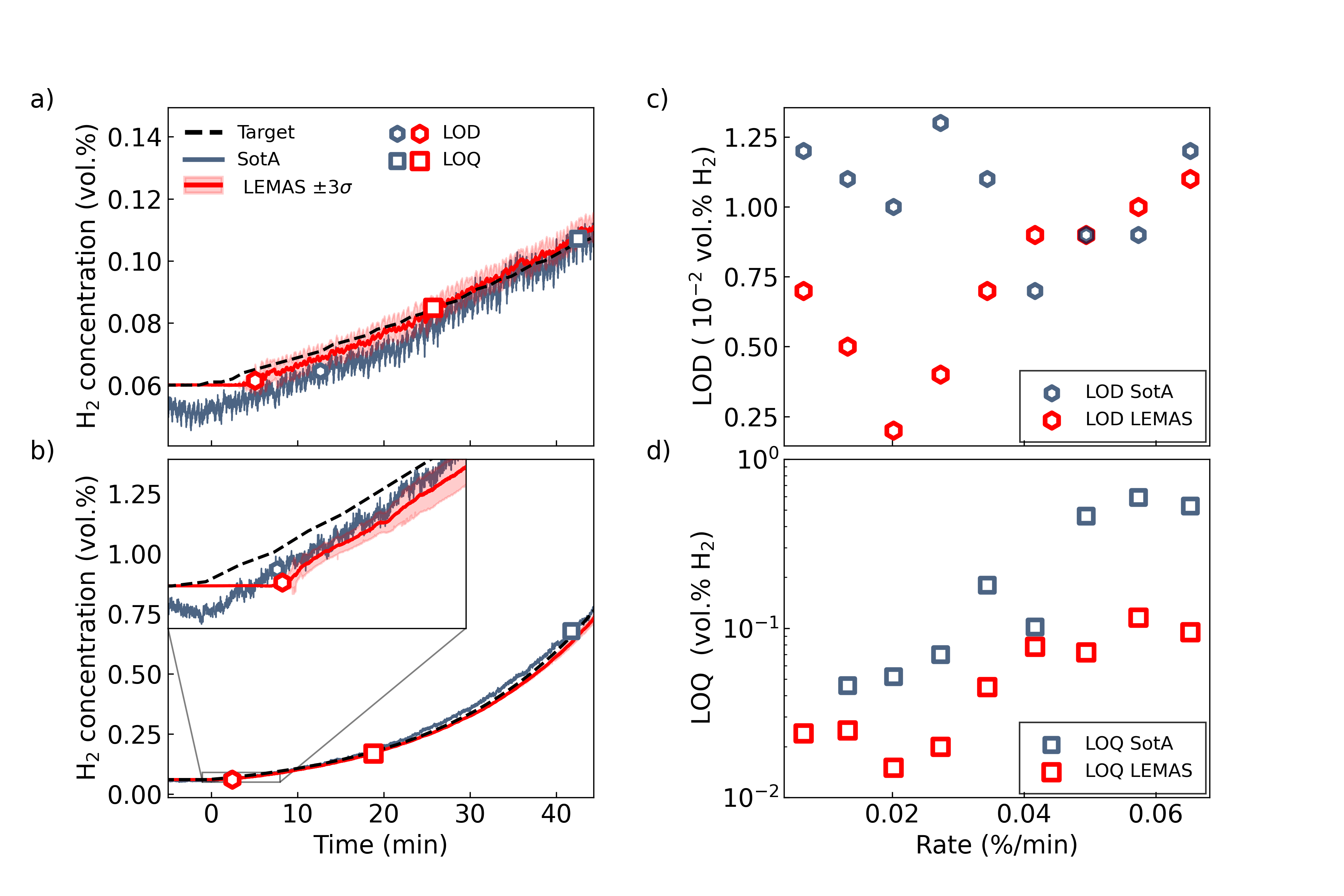}
    \caption{
        \textbf{Small leak detection and quantification.}
        (a-b) Time evolution of the \ce{H2} concentration obtained using the \acrshort{sota} centroid analysis and \gls{lemas}, respectively, for exponentially increasing \ce{H2} concentration in inert Ar environment for leak rates of (a) \qty{1.32e-3}{\hydrogenpercent\per\minute} and (b) \qty{5.73e-2}{\hydrogenpercent\per\minute}.
        The correspondingly obtained \acrfull{lod} and \acrfull{loq} are also indicated.
        (c-d) \acrshort{lod} and \acrshort{loq} as a function of leak rate.
        Note that at the smallest leak rate the \acrshort{lod} is not reached for the \acrshort{sota} centroid analysis.
    }
    \label{fig:leak_detection}
\end{figure*}

Finally, we note that for some pulses (illustrated by 0.08 and \qty{0.10}{\hydrogenpercent} in \autoref{fig:stepwise_accuracy}a) the predicted \ce{H2} concentration is overestimated for a brief interval past $t_{90}^\text{LEMAS}$, before the mean prediction settles around the target value.
Consequently, $t_s^\text{LEMAS}$ is larger than $t_{90}^\text{LEMAS}$.
Conversely, in other cases (illustrated by the pulse to \qty{0.06}{\hydrogenpercent}) the \ce{H2} concentration is underestimated until $t_{90}^\text{LEMAS}$, at which point the uncertainty has also been reduced.
As a result, $t_s^\text{LEMAS}$ equals $t_{90}^\text{LEMAS}$ in this (and similar) cases.
To further examine the overestimations we analyze the relative error of the predicted response, starting from $t_{90}^\text{LEMAS}$ (\autoref{fig:stepwise_accuracy}b).
Overall the relative error tends to be larger for lower \ce{H2} concentration pulses since there is a transient overestimation in the \gls{lemas} prediction.
Consequently, $t_s^\text{LEMAS}$ is generally larger for lower \ce{H2} concentrations (\autoref{fig:stepwise_accuracy}c).
This is likely the consequence of the early predictions being more affected by measurement noise, since lower \ce{H2} concentrations are associated with slower absorption kinetics and smaller changes in the extinction spectrum.
An important implication of these initial over-estimations is that the accuracy of the sensor initial response can be compromised if one relies on a single model, providing further evidence for the benefit of using an ensemble model, as we do with \gls{lemas}.

\paragraph{Improving sensor response to a simulated small and slow leak in inert Ar environment.} 
In a practical application in inert gas environment, \ce{H2} sensors are not only required for the rapid detection of large leaks with fast and essentially instantaneous increase of \ce{H2} concentration, but also in scenarios where a small leak will lead to a slow increase in \ce{H2} concentration in an enclosed environment over time.
Technically, this translates into the challenge of being able to as quickly as possible discern a tiny sensor signal from noise.
To address this scenario in the \gls{lemas} framework, we define the \gls{lod} of a sensor as the minimal amount of \ce{H2} required for the mean \ce{H2} prediction to change by more than three times the standard deviation of the \ce{H2} prediction at a baseline where the \ce{H2} concentration is kept constant.
In other words, the smallest \ce{H2} required to discern (but not quantify) the presence of hydrogen gas with a confidence of $3\sigma$. 
Furthermore, we define the \gls{loq} as the minimal amount of \ce{H2} for which the mean relative error of the \ce{H2} prediction is less than 5 \%.
The mean absolute relative error $\frac{1}{N} \sum_{t=1}^{N} \left| \frac{P_t - T_t}{T_t} \right|$ is calculated over a time window of \qty{2}{\minute}, where $N$ is the total number of time steps across the window, $P_t$ is the predicted \ce{H2} value for time step $t$ and $T_t$ is the true \ce{H2} value for time-step $t$.

We then assess the sensor response to a first scenario with a very small exponential leak rate of \qty{1.32e-3}{\hydrogenpercent\per\minute} using both the centroid-shift based \gls{sota} analysis and \gls{lemas} (\autoref{fig:leak_detection}a).
This analysis reveals that the \gls{lemas}-predicted \ce{H2} concentrations contain considerably less noise compared to the \gls{sota} analysis.
In the \gls{sota} analysis the noise in the \ce{H2} signal is initially comparable to the change in \ce{H2} concentration. 
As a result, the \gls{lod} is reached faster, i.e., at lower \ce{H2} concentrations, for \gls{lemas} analysis due to its ability to faster discern changes in the \ce{H2} concentration.
We also note that the \gls{lemas} analysis generally delivers a more accurate response, where the predicted \ce{H2} concentration values are closer to the target values, resulting in the \gls{lod} being reached earlier for \gls{lemas}. 

We also perform a similar analysis of a second scenario with a higher exponential leak of rate \qty{5.73e-2}{\hydrogenpercent\per\minute} (\autoref{fig:leak_detection}b; see also \autoref{sfig:all_leaks} for the \gls{sota} and \gls{lemas}-predicted \ce{H2} concentrations for all exponential leak rates).
Consistent with the previous analysis, \gls{lemas} demonstrates a 
significantly lower \gls{loq}, attributed to its overall higher accuracy.
However, despite \gls{lemas} exhibiting less noise than the \gls{sota} analysis the \gls{lod} of the two approaches are very similar.
This can be understood by noting that, in contrast to the previous case, here the fluctuations in the prediction of the \gls{sota} analysis are much smaller than the change in \ce{H2} due to the higher exponential leak rate.

These results are further corroborated when comparing the \gls{lod} and \gls{loq} obtained by the \gls{sota} and \gls{lemas} analysis methods, respectively, across a range of  exponential leak rates from \qtyrange[range-phrase=~to~]{6.52e-3}{6.52e-2}{\hydrogenpercent\per\minute} (\autoref{fig:leak_detection}c-d).
From \autoref{fig:leak_detection}c we identify that for exponential leak rates at \qty{3.44e-2}{\hydrogenpercent\per\minute} and below \gls{lemas} has a significantly lower \gls{lod} than \gls{sota}.
At larger exponential leak rates, the \gls{lod} for the \gls{lemas} and \gls{sota} analysis becomes approximately equal.
This occurs because, at larger rates, the change in \ce{H2} is sufficiently large, such that the initial change in the sensor signal is much larger than the intrinsic noise.
Consequently, the ability of \gls{lemas} to discern small signals from noise does not significantly contribute to decreasing the \gls{lod}.
At lower leak rates, however, \gls{lemas} indeed makes it possible to extract a discernible signal earlier, at lower leaked concentrations, thereby significantly increasing the time window from triggered sensor response to the leak having reached the flammability limit of \qty{4}{\hydrogenpercent}.
Finally, in \autoref{fig:leak_detection}d, we see that \gls{lemas} has a lower \gls{lod} for all investigated leak rates, which is a consequence of the higher accuracy obtained through \gls{lemas} (see \autoref{snote:sota-analysis}).
In summary, these results underscore on one hand the effectiveness of \gls{lemas} in detecting small and slow leaks earlier, as it consistently achieves a lower \gls{lod} than the \gls{sota} analysis at small rates.
On the other hand, they demonstrate that \gls{lemas} consistently outperforms the \gls{sota} analysis in terms of leak quantification, as reflected in its lower \gls{loq} across all rates.

\section{Conclusions}\label{sect:conclusions}

In this work, we have leveraged plasmonic \ce{H2} sensors with deep learning to address a crucial challenge in \ce{H2} sensing: the need for faster \ce{H2} detection in technically relevant conditions.
We have focused on the so far relatively unexplored but technically important application area of inert gas environments planned to be used to encapsulate/enclose large scale \ce{H2} installations, such as fuel pipes on \ce{H2}-powered airplanes or entire engine rooms on fuel-cell powered ships, to avoid the formation of flammable air-\ce{H2} mixtures. 
For this application area of \ce{H2} detection, hydride-forming plasmonic sensors are particularly well-suited, since they do not require molecular oxygen for their operation --- in contrast to, e.g., catalytic or thermal \ce{H2} sensors, which are most commonly used today.

From the deep learning perspective, we have developed \gls{lemas}, short for \acrlong{lemas}, which accelerates sensor response by learning the relationship between the time dependence of the full spectral response of the plasmonic sensor and the \ce{H2} concentration to predict the final sensor response before it is reached physically, while simultaneously assessing uncertainty in the model predictions through model ensembles.
To obtain accurate models for the ensemble and mitigate artifacts from measurement noise, drift, and variations between different measurements, we found it crucial to use a sufficiently long time series and combine the result of several different pre-processing approaches, such as wavelength dependent min-max normalization, standard normal variate standardization, global min-max normalization, and level scaling.

In summary, our results demonstrate the ability of deep learning concepts to significantly accelerate sensor response times and to enhance quantification of \ce{H2} leaks by enabling their detection earlier and therefore at lower leaked concentrations.
This, in turn, is important from a practical application perspective, since it provides a longer time window for the implementation of appropriate measures for handling the leak. 
Specifically, we have demonstrated (i) that \gls{lemas} is able to accelerate the response time of a plasmonic \ce{H2} sensor by a factor of 20 to 40 in an inert gas environment for \ce{H2} concentrations of \SI{0.1}{\hydrogenpercent} and below, and (ii) that it effectively eliminates the intrinsic \ce{H2} concentration dependence of metal hydride-based sensors. 
This is an important result because the characteristic increase of sensor response time for decreasing \ce{H2} concentrations constitutes one of the long-standing unresolved limitations in \ce{H2} sensor technology.

Taken together, our findings underscore the significant potential of deep learning for overcoming current limitations in \ce{H2} sensor performance, such as slow response in technologically relevant sensing environments like the inert gas environment investigated here.
Furthermore, while our investigation primarily focused on plasmonic \ce{H2} sensors, it is important to emphasize that the insights gained from this study hold broader implications for sensor technology in general as they illustrate a relatively unexplored generic concept for accelerating sensors by means of deep learning.

\appendix
\section{Methods}
\label{sect:methods}

\paragraph{Hydrogen sensing experiments.}
\label{sect:method-measurements}
The measurements were conducted in a custom-built reactor chamber that is comprised of a customized DN 16 CF spacer flange (Pfeifer Vacuum), equipped with a gas in- and outlet, and two fused-silica viewports (1.33” CF Flage, Accu-Glass).
The effective chamber volume is ca. \qty{1.5}{\milli\liter}.
The gas flow rates were controlled by mass flow controllers (El-Flow Select series, Bronkhorst High-Tech) (\autoref{sfig:minireactor}).
The sample inside the chamber was illuminated using an unpolarized halogen white light source (AvaLight-HAL, Avantes) and an optical fiber equipped with a collimating lens.
The transmitted light was collected and analyzed using a fiber-coupled fixed-grating spectrometer (SensLine AvaSpec-HS1024TEC, Avantes).
The temperature was controlled with a heating coil wrapped around the chamber and a temperature controller (Eurotherm 3216) in a feedback loop manner, where the sample surface temperature inside the chamber was continuously used as input.

All measurements were performed at \qty{30}{\celsius} in Argon background, with a constant gas flow of \qty[per-mode=repeated-symbol]{300}{\milli\liter\per\minute}.
The hydrogen concentration in all of the following measurements was in the range of \qtyrange[range-phrase=~--~]{0.06}{1.97}{\hydrogenpercent} (detailed description of the different pulse schemes as found in \autoref{snote:protocols}).
The sampling frequency of the spectrometer was set to \qty{3}{\hertz}.

\paragraph{SotA analysis.}
\label{sect:method-sota}
In this work, we used the centroid position as a spectral descriptor. 
The centroid position is defined as
$\lambda_c = \sum_\lambda \lambda I(\lambda)/\sum_\lambda I(\lambda)$,
where $\lambda$ is the wavelength in \qty{}{\nano\meter} and $I(\lambda)$ is the intensity at wavelength $\lambda$.
To enable comparison between the \gls{sota} analysis and \gls{lemas} on the test measurements we fit a calibration function, using the measured \ce{H2} concentration in the training measurements, as 
\begin{equation}
    \ce{H2}(\Delta\lambda_c) = a \Delta\lambda_c^b,
    \label{eq:sota_fit}
\end{equation}
where $\Delta \lambda_c$ is the change centroid position, taken from the smallest centroid position in each measurement. 
The values of the parameters $a$ and $b$ are determined by minimizing the mean absolute percentage error between the measured \ce{H2} and \ce{H2}$(\lambda_c)$.
We fit two different calibration functions, one for the data consisting of of step-wise increase/decrease of \ce{H2} and one for the data consisting of of linear/exponential increase/decrease of \ce{H2} (see \autoref{snote:sota-analysis} for details).

\paragraph{Data pre-processing.}
\label{sect:method-data-pre-proc}
Before the data was fed into the deep learning model it was pre-processed using four different methods, and the concatenation of these methods was fed as input to the deep learning model.
Each measurement was pre-processed individually by using the initial sequence of 5 pulses of \qty{1.97}{\hydrogenpercent}, for each measurement, to estimate the minimum/maximum/and mean intensity.
The scaling methods were (i) wavelength dependent min-max normalization: for each wavelength subtracting the estimated minimum intensity at the corresponding wavelength and dividing by the difference between the estimated maximum intensity of all wavelengths and the estimated minimum intensity measured at the specific wavelength, (ii) standard normal variate standardization: scaling each spectrum using its mean and standard deviation, (iii) global min-max normalization: subtracting the estimated minimum intensity and dividing by the difference between the estimated maximum and minimum intensity in and (iv) level scaling: subtracting and dividing each spectrum in the measurement by the estimated mean intensity.

\paragraph{Deep learning model.}
\label{sect:method-model}
The deep learning architecture that was used in this work was a \acrfull{lstr} \cite{Xu2021} which operates as illustrated in \autoref{sfig:SI_detailed_lstr}. 
Each temporal feature consisting of the concatenation is linearly mapped to a vector of size $d_{model} = 256$.
Subsequently, positional encoding is added and the data is split into a short-term memory and long-term memory.Here, we down-sample the long-term memory using a stride of 4.
Firstly, the long-term memory undergoes a two stage memory compression through the \gls{lstr} encoder, using a set of learnable token embeddings of dimensions $d_{model} \times n_1$ and $d_{model} \times n_0$. 
Here, we used $n_0 = 8$ and $n_1 =4$ and the
encoder consisted of 4 transformer decoder units.
Secondly, the \gls{lstr} decoder extracts relevant temporal features in the short-term memory, while also querying the encoded long-term memory to retrieve useful information from the history of the sensor.
Here, the decoder consisted of 8 transformer decoder units.
The extracted temporal features are then passed through $n_{mlp} = 8$ \gls{mlp} layers of dimension $ d_{mlp} = 512$ to obtain \ce{H2} concentration predictions.
Here all the transformer decoder units performed multi head attention as in with $h=8$  heads and the $d_{k} = d_{q} = d_{v}  = d_{model}/h = 32$, and the dimension of the \gls{mlp} inside the transformer decoder units was $d_{ff} = 512$.
Furthermore, in the \gls{lstr} encoder, masked multi-head attention was performed such that during training the \ce{H2} concentration corresponding to each time step in the short term memory could be used for supervision during training.

\paragraph{Ensembles}
\label{sect:method-ensembles}
The constructed ensembles comprised ten models, each varying in the lengths of short-term and long-term memory.
This variation was designed to induce diversity in the predictive capabilities of the models within the ensemble.
Specifically, two models were designated for each combination of long and short-term memory lengths, while ensuring a consistent total input sequence length across all models.

For the ensemble tailored for leak detection, the preceding \num{4000} time steps constituted the input.
By contrast, the ensemble model, which was aimed at minimizing response time, utilized the previous \num{600} time steps as input.
For both ensembles, the selected lengths for short-term memory were 20, 40, 60, 80, and 100 time steps, respectively.
This approach was adopted to enhance the generalization capabilities the ensemble.
It is important to note that apart from the variation in memory lengths, all models shared identical hyperparameters (see \autoref{stable:hyperparam}).

To make the prediction of the ensemble more robust to potential outliers, we only included predictions that fall between the first and third percentile to compute the ensemble prediction and uncertainty.

\paragraph{Deep learning training}
\label{sect:method-deep-learning-training}
The models were implemented using \textsc{TensorFlow} \cite{tensorflow2015-whitepaper} and were trained for \num{100} epochs on Nvidia A100 graphical processing units using the AdamW \cite{loshchilov2019decoupled} optimizer with weight decay \num{5e-5}, a batch size of \num{128}, and mean-absolute-error loss.
The learning rate was increased linearly from zero to \num{5e-5} during the first \num{15} epochs then decaying to zero following a cosine curve.
To analyze the impact of different pre-processing methods we used the first half of the data from measurement \autoref{sfig:step protocols}a and \autoref{sfig:step protocols}c as training data and the other half as validation data (see \autoref{snote:data-pre-processing}).
For the first case of a large simulated leak characterized by a rapid step-wise increase of \ce{H2} concentration in the sensor surroundings, we trained \gls{lemas} using the data from measurement \autoref{sfig:step protocols}a and \autoref{sfig:step protocols}c as training data and the data from measurement \autoref{sfig:step protocols}b as test data.
For the second case of a simulated small slow leak, we trained \gls{lemas} the data from measurement \autoref{sfig:leak protocols}a as training data and data from measurement \autoref{sfig:leak protocols}b as test data.
Furthermore, during the training phase, each model in the ensemble was exposed to a distinct subset of the training data, comprising a random 90\% of the total dataset.

%----------------------------------------------------------%
\paragraph{Sample fabrication.}
\label{sect:method-sample}
Quasi-random \ce{PdAu} alloy (nominal 70:30 at. \%) nanodisk arrays with \qty{210}{\nano\meter} average disk diameter and \qty{25}{\nano\meter} height, were fabricated using Hole-Mask Colloidal Lithography (HCL) \cite{Fredriksson2007}.
The metals were deposited layer-by-layer via electron beam evaporation, onto \qtyproduct{1x1}{\centi\meter} fused silica substrates (Siegert Wafer GmbH).
Subsequent annealing was performed at \qty{500}{\celsius} for \qty{18}{\hour} under a flow of \qty{4}{\hydrogenpercent} in Ar to induce alloy formation.
A more detailed description of the nanofabrication procedure can be found in our earlier work \cite{NugIanWag16}.

\section{Data availability}

\section*{Author contributions}

\section*{Competing interests}
C.~L. is co-founder and scientific advisor at Insplorion AB who markets plasmonic hydrogen sensors.

\section*{Acknowledgments}
This work was funded by the Vinnova project 2021-02760, the Swedish Research Council (grant numbers 2018-06482, 2020-04935, 2021-05072), the Swedish Energy Agency (grant No. 45410-1), the Area of Advanced Nano at Chalmers, and the Competence Centre TechForH2.
The Competence Centre TechForH2 is hosted by Chalmers University of Technology and is financially supported by the Swedish Energy Agency (P2021-90268) and the member companies Volvo, Scania, Siemens Energy, GKN Aerospace, PowerCell, Oxeon, RISE, Stena Rederier AB, Johnsson Matthey and Insplorion.
The computations were enabled by resources provided by the National Academic Infrastructure for Supercomputing in Sweden (NAISS) at C3SE partially funded by the Swedish Research Council through grant agreement no.\ 2022-06725.
This work was performed in part at Myfab Chalmers and the Chalmers Materials Analysis Laboratory (CMAL).


\begin{thebibliography}{34}%
\makeatletter
\providecommand \@ifxundefined [1]{%
 \@ifx{#1\undefined}
}%
\providecommand \@ifnum [1]{%
 \ifnum #1\expandafter \@firstoftwo
 \else \expandafter \@secondoftwo
 \fi
}%
\providecommand \@ifx [1]{%
 \ifx #1\expandafter \@firstoftwo
 \else \expandafter \@secondoftwo
 \fi
}%
\providecommand \natexlab [1]{#1}%
\providecommand \enquote  [1]{``#1''}%
\providecommand \bibnamefont  [1]{#1}%
\providecommand \bibfnamefont [1]{#1}%
\providecommand \citenamefont [1]{#1}%
\providecommand \href@noop [0]{\@secondoftwo}%
\providecommand \href [0]{\begingroup \@sanitize@url \@href}%
\providecommand \@href[1]{\@@startlink{#1}\@@href}%
\providecommand \@@href[1]{\endgroup#1\@@endlink}%
\providecommand \@sanitize@url [0]{\catcode `\\12\catcode `\$12\catcode
  `\&12\catcode `\#12\catcode `\^12\catcode `\_12\catcode `\%12\relax}%
\providecommand \@@startlink[1]{}%
\providecommand \@@endlink[0]{}%
\providecommand \url  [0]{\begingroup\@sanitize@url \@url }%
\providecommand \@url [1]{\endgroup\@href {#1}{\urlprefix }}%
\providecommand \urlprefix  [0]{URL }%
\providecommand \Eprint [0]{\href }%
\providecommand \doibase [0]{https://doi.org/}%
\providecommand \selectlanguage [0]{\@gobble}%
\providecommand \bibinfo  [0]{\@secondoftwo}%
\providecommand \bibfield  [0]{\@secondoftwo}%
\providecommand \translation [1]{[#1]}%
\providecommand \BibitemOpen [0]{}%
\providecommand \bibitemStop [0]{}%
\providecommand \bibitemNoStop [0]{.\EOS\space}%
\providecommand \EOS [0]{\spacefactor3000\relax}%
\providecommand \BibitemShut  [1]{\csname bibitem#1\endcsname}%
\let\auto@bib@innerbib\@empty
%</preamble>
\bibitem [{\citenamefont {Zhang}\ \emph {et~al.}(2010)\citenamefont {Zhang},
  \citenamefont {Yuan}, \citenamefont {Song},\ and\ \citenamefont
  {Zheng}}]{ZHANG201087}%
  \BibitemOpen
  \bibfield  {author} {\bibinfo {author} {\bibfnamefont {M.}~\bibnamefont
  {Zhang}}, \bibinfo {author} {\bibfnamefont {Z.}~\bibnamefont {Yuan}},
  \bibinfo {author} {\bibfnamefont {J.}~\bibnamefont {Song}},\ and\ \bibinfo
  {author} {\bibfnamefont {C.}~\bibnamefont {Zheng}},\ }\href
  {https://doi.org/10.1016/j.snb.2010.05.001} {\bibfield  {journal} {\bibinfo
  {journal} {Sensors and Actuators B: Chemical}\ }\textbf {\bibinfo {volume}
  {148}},\ \bibinfo {pages} {87} (\bibinfo {year} {2010})}\BibitemShut
  {NoStop}%
\bibitem [{\citenamefont {Ribet}\ \emph {et~al.}(2018)\citenamefont {Ribet},
  \citenamefont {Stemme},\ and\ \citenamefont {Roxhed}}]{ribet2018real}%
  \BibitemOpen
  \bibfield  {author} {\bibinfo {author} {\bibfnamefont {F.}~\bibnamefont
  {Ribet}}, \bibinfo {author} {\bibfnamefont {G.}~\bibnamefont {Stemme}},\ and\
  \bibinfo {author} {\bibfnamefont {N.}~\bibnamefont {Roxhed}},\ }\href
  {https://doi.org/10.1007/s10544-018-0349-6} {\bibfield  {journal} {\bibinfo
  {journal} {Biomedical microdevices}\ }\textbf {\bibinfo {volume} {20}},\
  \bibinfo {pages} {1} (\bibinfo {year} {2018})}\BibitemShut {NoStop}%
\bibitem [{\citenamefont {Rolinger}\ \emph {et~al.}(2020)\citenamefont
  {Rolinger}, \citenamefont {R{\"u}dt},\ and\ \citenamefont
  {Hubbuch}}]{RolRudHub20}%
  \BibitemOpen
  \bibfield  {author} {\bibinfo {author} {\bibfnamefont {L.}~\bibnamefont
  {Rolinger}}, \bibinfo {author} {\bibfnamefont {M.}~\bibnamefont {R{\"u}dt}},\
  and\ \bibinfo {author} {\bibfnamefont {J.}~\bibnamefont {Hubbuch}},\ }\href
  {https://doi.org/10.1007/s00216-020-02407-z} {\bibfield  {journal} {\bibinfo
  {journal} {Analytical and Bioanalytical Chemistry}\ }\textbf {\bibinfo
  {volume} {412}},\ \bibinfo {pages} {2047} (\bibinfo {year}
  {2020})}\BibitemShut {NoStop}%
\bibitem [{\citenamefont {Poghossian}\ \emph {et~al.}(2019)\citenamefont
  {Poghossian}, \citenamefont {Geissler},\ and\ \citenamefont
  {Schöning}}]{POGHOSSIAN2019111272}%
  \BibitemOpen
  \bibfield  {author} {\bibinfo {author} {\bibfnamefont {A.}~\bibnamefont
  {Poghossian}}, \bibinfo {author} {\bibfnamefont {H.}~\bibnamefont
  {Geissler}},\ and\ \bibinfo {author} {\bibfnamefont {M.~J.}\ \bibnamefont
  {Schöning}},\ }\href {https://doi.org/10.1016/j.bios.2019.04.040} {\bibfield
   {journal} {\bibinfo  {journal} {Biosensors and Bioelectronics}\ }\textbf
  {\bibinfo {volume} {140}},\ \bibinfo {pages} {111272} (\bibinfo {year}
  {2019})}\BibitemShut {NoStop}%
\bibitem [{\citenamefont {Darmadi}\ \emph {et~al.}(2020)\citenamefont
  {Darmadi}, \citenamefont {Nugroho},\ and\ \citenamefont
  {Langhammer}}]{DarNugLan20}%
  \BibitemOpen
  \bibfield  {author} {\bibinfo {author} {\bibfnamefont {I.}~\bibnamefont
  {Darmadi}}, \bibinfo {author} {\bibfnamefont {F.~A.~A.}\ \bibnamefont
  {Nugroho}},\ and\ \bibinfo {author} {\bibfnamefont {C.}~\bibnamefont
  {Langhammer}},\ }\href {https://doi.org/10.1021/acssensors.0c02019}
  {\bibfield  {journal} {\bibinfo  {journal} {ACS Sensors}\ }\textbf {\bibinfo
  {volume} {5}},\ \bibinfo {pages} {3306} (\bibinfo {year} {2020})}\BibitemShut
  {NoStop}%
\bibitem [{doe(2015)}]{doe-targets}%
  \BibitemOpen
  \href
  {https://www.energy.gov/sites/default/files/2015/06/f23/fcto_myrdd_safety_codes.pdf}
  {\bibinfo {title} {U.s. department of energy; 2015 safety, codes and
  standards section}} (\bibinfo {year} {2015})\BibitemShut {NoStop}%
\bibitem [{\citenamefont {Nugroho}\ \emph {et~al.}(2019)\citenamefont
  {Nugroho}, \citenamefont {Darmadi}, \citenamefont {Cusinato}, \citenamefont
  {Susarrey-Arce}, \citenamefont {Schreuders}, \citenamefont {Bannenberg},
  \citenamefont {da~Silva~Fanta}, \citenamefont {Kadkhodazadeh}, \citenamefont
  {Wagner}, \citenamefont {Antosiewicz} \emph {et~al.}}]{NugDarCus19}%
  \BibitemOpen
  \bibfield  {author} {\bibinfo {author} {\bibfnamefont {F.~A.}\ \bibnamefont
  {Nugroho}}, \bibinfo {author} {\bibfnamefont {I.}~\bibnamefont {Darmadi}},
  \bibinfo {author} {\bibfnamefont {L.}~\bibnamefont {Cusinato}}, \bibinfo
  {author} {\bibfnamefont {A.}~\bibnamefont {Susarrey-Arce}}, \bibinfo {author}
  {\bibfnamefont {H.}~\bibnamefont {Schreuders}}, \bibinfo {author}
  {\bibfnamefont {L.~J.}\ \bibnamefont {Bannenberg}}, \bibinfo {author}
  {\bibfnamefont {A.~B.}\ \bibnamefont {da~Silva~Fanta}}, \bibinfo {author}
  {\bibfnamefont {S.}~\bibnamefont {Kadkhodazadeh}}, \bibinfo {author}
  {\bibfnamefont {J.~B.}\ \bibnamefont {Wagner}}, \bibinfo {author}
  {\bibfnamefont {T.~J.}\ \bibnamefont {Antosiewicz}}, \emph {et~al.},\ }\href
  {https://doi.org/10.1038/s41563-019-0325-4} {\bibfield  {journal} {\bibinfo
  {journal} {Nature Materials}\ }\textbf {\bibinfo {volume} {18}},\ \bibinfo
  {pages} {489} (\bibinfo {year} {2019})}\BibitemShut {NoStop}%
\bibitem [{\citenamefont {Luong}\ \emph {et~al.}(2021)\citenamefont {Luong},
  \citenamefont {Pham}, \citenamefont {Guin}, \citenamefont {Madhogaria},
  \citenamefont {Phan}, \citenamefont {Larsen},\ and\ \citenamefont
  {Nguyen}}]{luong2021sub}%
  \BibitemOpen
  \bibfield  {author} {\bibinfo {author} {\bibfnamefont {H.~M.}\ \bibnamefont
  {Luong}}, \bibinfo {author} {\bibfnamefont {M.~T.}\ \bibnamefont {Pham}},
  \bibinfo {author} {\bibfnamefont {T.}~\bibnamefont {Guin}}, \bibinfo {author}
  {\bibfnamefont {R.~P.}\ \bibnamefont {Madhogaria}}, \bibinfo {author}
  {\bibfnamefont {M.-H.}\ \bibnamefont {Phan}}, \bibinfo {author}
  {\bibfnamefont {G.~K.}\ \bibnamefont {Larsen}},\ and\ \bibinfo {author}
  {\bibfnamefont {T.~D.}\ \bibnamefont {Nguyen}},\ }\href
  {https://doi.org/10.1038/s41467-021-22697-w} {\bibfield  {journal} {\bibinfo
  {journal} {Nature communications}\ }\textbf {\bibinfo {volume} {12}},\
  \bibinfo {pages} {2414} (\bibinfo {year} {2021})}\BibitemShut {NoStop}%
\bibitem [{\citenamefont {Bannenberg}\ \emph {et~al.}(2021)\citenamefont
  {Bannenberg}, \citenamefont {Schreuders},\ and\ \citenamefont
  {Dam}}]{BanSch21}%
  \BibitemOpen
  \bibfield  {author} {\bibinfo {author} {\bibfnamefont {L.}~\bibnamefont
  {Bannenberg}}, \bibinfo {author} {\bibfnamefont {H.}~\bibnamefont
  {Schreuders}},\ and\ \bibinfo {author} {\bibfnamefont {B.}~\bibnamefont
  {Dam}},\ }\href {https://doi.org/10.1002/adfm.202010483} {\bibfield
  {journal} {\bibinfo  {journal} {Advanced Functional Materials}\ }\textbf
  {\bibinfo {volume} {31}},\ \bibinfo {pages} {2010483} (\bibinfo {year}
  {2021})}\BibitemShut {NoStop}%
\bibitem [{\citenamefont {Darmadi}\ \emph {et~al.}(2019)\citenamefont
  {Darmadi}, \citenamefont {Nugroho}, \citenamefont {Kadkhodazadeh},
  \citenamefont {Wagner},\ and\ \citenamefont {Langhammer}}]{DarNugKad19}%
  \BibitemOpen
  \bibfield  {author} {\bibinfo {author} {\bibfnamefont {I.}~\bibnamefont
  {Darmadi}}, \bibinfo {author} {\bibfnamefont {F.~A.~A.}\ \bibnamefont
  {Nugroho}}, \bibinfo {author} {\bibfnamefont {S.}~\bibnamefont
  {Kadkhodazadeh}}, \bibinfo {author} {\bibfnamefont {J.~B.}\ \bibnamefont
  {Wagner}},\ and\ \bibinfo {author} {\bibfnamefont {C.}~\bibnamefont
  {Langhammer}},\ }\href {https://doi.org/10.1021/acssensors.9b00610}
  {\bibfield  {journal} {\bibinfo  {journal} {ACS Sensors}\ }\textbf {\bibinfo
  {volume} {4}},\ \bibinfo {pages} {1424} (\bibinfo {year} {2019})}\BibitemShut
  {NoStop}%
\bibitem [{\citenamefont {Alenezy}\ \emph {et~al.}(2020)\citenamefont
  {Alenezy}, \citenamefont {Sabri}, \citenamefont {Kandjani}, \citenamefont
  {Korcoban}, \citenamefont {Abdul Haroon~Rashid}, \citenamefont {Ippolito},\
  and\ \citenamefont {Bhargava}}]{AlEbSa2020}%
  \BibitemOpen
  \bibfield  {author} {\bibinfo {author} {\bibfnamefont {E.~K.}\ \bibnamefont
  {Alenezy}}, \bibinfo {author} {\bibfnamefont {Y.~M.}\ \bibnamefont {Sabri}},
  \bibinfo {author} {\bibfnamefont {A.~E.}\ \bibnamefont {Kandjani}}, \bibinfo
  {author} {\bibfnamefont {D.}~\bibnamefont {Korcoban}}, \bibinfo {author}
  {\bibfnamefont {S.~S.~A.}\ \bibnamefont {Abdul Haroon~Rashid}}, \bibinfo
  {author} {\bibfnamefont {S.~J.}\ \bibnamefont {Ippolito}},\ and\ \bibinfo
  {author} {\bibfnamefont {S.~K.}\ \bibnamefont {Bhargava}},\ }\href
  {https://doi.org/10.1021/acssensors.0c01387} {\bibfield  {journal} {\bibinfo
  {journal} {ACS Sensors}\ }\textbf {\bibinfo {volume} {5}},\ \bibinfo {pages}
  {3902} (\bibinfo {year} {2020})}\BibitemShut {NoStop}%
\bibitem [{\citenamefont {Kabcum}\ \emph {et~al.}(2016)\citenamefont {Kabcum},
  \citenamefont {Channei}, \citenamefont {Tuantranont}, \citenamefont
  {Wisitsoraat}, \citenamefont {Liewhiran},\ and\ \citenamefont
  {Phanichphant}}]{KABCUM201676}%
  \BibitemOpen
  \bibfield  {author} {\bibinfo {author} {\bibfnamefont {S.}~\bibnamefont
  {Kabcum}}, \bibinfo {author} {\bibfnamefont {D.}~\bibnamefont {Channei}},
  \bibinfo {author} {\bibfnamefont {A.}~\bibnamefont {Tuantranont}}, \bibinfo
  {author} {\bibfnamefont {A.}~\bibnamefont {Wisitsoraat}}, \bibinfo {author}
  {\bibfnamefont {C.}~\bibnamefont {Liewhiran}},\ and\ \bibinfo {author}
  {\bibfnamefont {S.}~\bibnamefont {Phanichphant}},\ }\href
  {https://doi.org/10.1016/j.snb.2015.11.120} {\bibfield  {journal} {\bibinfo
  {journal} {Sensors and Actuators B: Chemical}\ }\textbf {\bibinfo {volume}
  {226}},\ \bibinfo {pages} {76} (\bibinfo {year} {2016})}\BibitemShut
  {NoStop}%
\bibitem [{\citenamefont {Hassan}\ \emph {et~al.}(2016)\citenamefont {Hassan},
  \citenamefont {{Iftekhar Uddin}},\ and\ \citenamefont
  {Chung}}]{HASSAN2016435}%
  \BibitemOpen
  \bibfield  {author} {\bibinfo {author} {\bibfnamefont {K.}~\bibnamefont
  {Hassan}}, \bibinfo {author} {\bibfnamefont {A.}~\bibnamefont {{Iftekhar
  Uddin}}},\ and\ \bibinfo {author} {\bibfnamefont {G.-S.}\ \bibnamefont
  {Chung}},\ }\href {https://doi.org/10.1016/j.snb.2016.05.013} {\bibfield
  {journal} {\bibinfo  {journal} {Sensors and Actuators B: Chemical}\ }\textbf
  {\bibinfo {volume} {234}},\ \bibinfo {pages} {435} (\bibinfo {year}
  {2016})}\BibitemShut {NoStop}%
\bibitem [{\citenamefont {Losurdo}\ \emph {et~al.}(2021)\citenamefont
  {Losurdo}, \citenamefont {Guti{\'e}rrez}, \citenamefont {Suvorova},
  \citenamefont {Giangregorio}, \citenamefont {Rubanov}, \citenamefont
  {Brown},\ and\ \citenamefont {Moreno}}]{LoMaGu2021}%
  \BibitemOpen
  \bibfield  {author} {\bibinfo {author} {\bibfnamefont {M.}~\bibnamefont
  {Losurdo}}, \bibinfo {author} {\bibfnamefont {Y.}~\bibnamefont
  {Guti{\'e}rrez}}, \bibinfo {author} {\bibfnamefont {A.}~\bibnamefont
  {Suvorova}}, \bibinfo {author} {\bibfnamefont {M.~M.}\ \bibnamefont
  {Giangregorio}}, \bibinfo {author} {\bibfnamefont {S.}~\bibnamefont
  {Rubanov}}, \bibinfo {author} {\bibfnamefont {A.~S.}\ \bibnamefont {Brown}},\
  and\ \bibinfo {author} {\bibfnamefont {F.}~\bibnamefont {Moreno}},\
  }\href@noop {} {\bibfield  {journal} {\bibinfo  {journal} {Advanced
  Materials}\ }\textbf {\bibinfo {volume} {33}},\ \bibinfo {pages} {2100500}
  (\bibinfo {year} {2021})}\BibitemShut {NoStop}%
\bibitem [{\citenamefont {Huang}\ \emph {et~al.}(2023)\citenamefont {Huang},
  \citenamefont {Croy}, \citenamefont {Ibarlucea},\ and\ \citenamefont
  {Cuniberti}}]{Huang2023}%
  \BibitemOpen
  \bibfield  {author} {\bibinfo {author} {\bibfnamefont {S.}~\bibnamefont
  {Huang}}, \bibinfo {author} {\bibfnamefont {A.}~\bibnamefont {Croy}},
  \bibinfo {author} {\bibfnamefont {B.}~\bibnamefont {Ibarlucea}},\ and\
  \bibinfo {author} {\bibfnamefont {G.}~\bibnamefont {Cuniberti}},\ }\bibinfo
  {title} {Machine learning-driven gas identification in gas sensors},\ in\
  \href {https://doi.org/10.1007/978-981-99-0393-1_2} {\emph {\bibinfo
  {booktitle} {Machine Learning for Advanced Functional Materials}}},\ \bibinfo
  {editor} {edited by\ \bibinfo {editor} {\bibfnamefont {N.}~\bibnamefont
  {Joshi}}, \bibinfo {editor} {\bibfnamefont {V.}~\bibnamefont {Kushvaha}},\
  and\ \bibinfo {editor} {\bibfnamefont {P.}~\bibnamefont {Madhushri}}}\
  (\bibinfo  {publisher} {Springer Nature Singapore},\ \bibinfo {address}
  {Singapore},\ \bibinfo {year} {2023})\ pp.\ \bibinfo {pages}
  {21--41}\BibitemShut {NoStop}%
\bibitem [{\citenamefont {Kang}\ \emph {et~al.}(2022)\citenamefont {Kang},
  \citenamefont {Cho}, \citenamefont {Park}, \citenamefont {Jeong},
  \citenamefont {Lee}, \citenamefont {Lee}, \citenamefont {Del Orbe~Henriquez},
  \citenamefont {Yoon},\ and\ \citenamefont {Park}}]{KanMinCho2022}%
  \BibitemOpen
  \bibfield  {author} {\bibinfo {author} {\bibfnamefont {M.}~\bibnamefont
  {Kang}}, \bibinfo {author} {\bibfnamefont {I.}~\bibnamefont {Cho}}, \bibinfo
  {author} {\bibfnamefont {J.}~\bibnamefont {Park}}, \bibinfo {author}
  {\bibfnamefont {J.}~\bibnamefont {Jeong}}, \bibinfo {author} {\bibfnamefont
  {K.}~\bibnamefont {Lee}}, \bibinfo {author} {\bibfnamefont {B.}~\bibnamefont
  {Lee}}, \bibinfo {author} {\bibfnamefont {D.}~\bibnamefont {Del
  Orbe~Henriquez}}, \bibinfo {author} {\bibfnamefont {K.}~\bibnamefont
  {Yoon}},\ and\ \bibinfo {author} {\bibfnamefont {I.}~\bibnamefont {Park}},\
  }\href {https://doi.org/10.1021/acssensors.1c01204} {\bibfield  {journal}
  {\bibinfo  {journal} {ACS Sensors}\ }\textbf {\bibinfo {volume} {7}},\
  \bibinfo {pages} {430} (\bibinfo {year} {2022})}\BibitemShut {NoStop}%
\bibitem [{\citenamefont {Ai}\ \emph {et~al.}(2022)\citenamefont {Ai},
  \citenamefont {Sun},\ and\ \citenamefont {Zhao}}]{AiBinSun}%
  \BibitemOpen
  \bibfield  {author} {\bibinfo {author} {\bibfnamefont {B.}~\bibnamefont
  {Ai}}, \bibinfo {author} {\bibfnamefont {Y.}~\bibnamefont {Sun}},\ and\
  \bibinfo {author} {\bibfnamefont {Y.}~\bibnamefont {Zhao}},\ }\href
  {https://doi.org/10.1002/smll.202107882} {\bibfield  {journal} {\bibinfo
  {journal} {Small}\ }\textbf {\bibinfo {volume} {18}},\ \bibinfo {pages}
  {2107882} (\bibinfo {year} {2022})}\BibitemShut {NoStop}%
\bibitem [{\citenamefont {Wadell}\ \emph {et~al.}(2014)\citenamefont {Wadell},
  \citenamefont {Syrenova},\ and\ \citenamefont {Langhammer}}]{WadSyrLan14}%
  \BibitemOpen
  \bibfield  {author} {\bibinfo {author} {\bibfnamefont {C.}~\bibnamefont
  {Wadell}}, \bibinfo {author} {\bibfnamefont {S.}~\bibnamefont {Syrenova}},\
  and\ \bibinfo {author} {\bibfnamefont {C.}~\bibnamefont {Langhammer}},\
  }\href {https://doi.org/10.1021/nn505804f} {\bibfield  {journal} {\bibinfo
  {journal} {ACS Nano}\ }\textbf {\bibinfo {volume} {8}},\ \bibinfo {pages}
  {11925} (\bibinfo {year} {2014})}\BibitemShut {NoStop}%
\bibitem [{\citenamefont {Dahlin}\ \emph {et~al.}(2006)\citenamefont {Dahlin},
  \citenamefont {Tegenfeldt},\ and\ \citenamefont {Höök}}]{DahTeg06}%
  \BibitemOpen
  \bibfield  {author} {\bibinfo {author} {\bibfnamefont {A.~B.}\ \bibnamefont
  {Dahlin}}, \bibinfo {author} {\bibfnamefont {J.~O.}\ \bibnamefont
  {Tegenfeldt}},\ and\ \bibinfo {author} {\bibfnamefont {F.}~\bibnamefont
  {Höök}},\ }\href {https://doi.org/10.1021/ac0601967} {\bibfield  {journal}
  {\bibinfo  {journal} {Analytical Chemistry}\ }\textbf {\bibinfo {volume}
  {78}},\ \bibinfo {pages} {4416} (\bibinfo {year} {2006})}\BibitemShut
  {NoStop}%
\bibitem [{\citenamefont {Xu}\ \emph {et~al.}(2021)\citenamefont {Xu},
  \citenamefont {Xiong}, \citenamefont {Chen}, \citenamefont {Li},
  \citenamefont {Xia}, \citenamefont {Tu},\ and\ \citenamefont
  {Soatto}}]{Xu2021}%
  \BibitemOpen
  \bibfield  {author} {\bibinfo {author} {\bibfnamefont {M.}~\bibnamefont
  {Xu}}, \bibinfo {author} {\bibfnamefont {Y.}~\bibnamefont {Xiong}}, \bibinfo
  {author} {\bibfnamefont {H.}~\bibnamefont {Chen}}, \bibinfo {author}
  {\bibfnamefont {X.}~\bibnamefont {Li}}, \bibinfo {author} {\bibfnamefont
  {W.}~\bibnamefont {Xia}}, \bibinfo {author} {\bibfnamefont {Z.}~\bibnamefont
  {Tu}},\ and\ \bibinfo {author} {\bibfnamefont {S.}~\bibnamefont {Soatto}},\
  }in\ \href
  {https://www.amazon.science/publications/long-short-term-transformer-for-online-action-detection}
  {\emph {\bibinfo {booktitle} {NeurIPS 2021}}}\ (\bibinfo {year}
  {2021})\BibitemShut {NoStop}%
\bibitem [{\citenamefont {Wadell}\ \emph {et~al.}(2015)\citenamefont {Wadell},
  \citenamefont {Nugroho}, \citenamefont {Lidstr\"om}, \citenamefont {Iandolo},
  \citenamefont {Wagner},\ and\ \citenamefont {Langhammer}}]{WadNugLid15}%
  \BibitemOpen
  \bibfield  {author} {\bibinfo {author} {\bibfnamefont {C.}~\bibnamefont
  {Wadell}}, \bibinfo {author} {\bibfnamefont {F.~A.~A.}\ \bibnamefont
  {Nugroho}}, \bibinfo {author} {\bibfnamefont {E.}~\bibnamefont {Lidstr\"om}},
  \bibinfo {author} {\bibfnamefont {B.}~\bibnamefont {Iandolo}}, \bibinfo
  {author} {\bibfnamefont {J.~B.}\ \bibnamefont {Wagner}},\ and\ \bibinfo
  {author} {\bibfnamefont {C.}~\bibnamefont {Langhammer}},\ }\href
  {https://doi.org/10.1021/acs.nanolett.5b01053} {\bibfield  {journal}
  {\bibinfo  {journal} {Nano Letters}\ }\textbf {\bibinfo {volume} {15}},\
  \bibinfo {pages} {3563} (\bibinfo {year} {2015})}\BibitemShut {NoStop}%
\bibitem [{\citenamefont {Bannenberg}\ \emph {et~al.}(2019)\citenamefont
  {Bannenberg}, \citenamefont {Nugroho}, \citenamefont {Schreuders},
  \citenamefont {Norder}, \citenamefont {Trinh}, \citenamefont {Steinke},
  \citenamefont {van Well}, \citenamefont {Langhammer},\ and\ \citenamefont
  {Dam}}]{BanNugSch19}%
  \BibitemOpen
  \bibfield  {author} {\bibinfo {author} {\bibfnamefont {L.~J.}\ \bibnamefont
  {Bannenberg}}, \bibinfo {author} {\bibfnamefont {F.~A.~A.}\ \bibnamefont
  {Nugroho}}, \bibinfo {author} {\bibfnamefont {H.}~\bibnamefont {Schreuders}},
  \bibinfo {author} {\bibfnamefont {B.}~\bibnamefont {Norder}}, \bibinfo
  {author} {\bibfnamefont {T.~T.}\ \bibnamefont {Trinh}}, \bibinfo {author}
  {\bibfnamefont {N.-J.}\ \bibnamefont {Steinke}}, \bibinfo {author}
  {\bibfnamefont {A.~A.}\ \bibnamefont {van Well}}, \bibinfo {author}
  {\bibfnamefont {C.}~\bibnamefont {Langhammer}},\ and\ \bibinfo {author}
  {\bibfnamefont {B.}~\bibnamefont {Dam}},\ }\href
  {https://doi.org/10.1021/acsami.8b22455} {\bibfield  {journal} {\bibinfo
  {journal} {ACS Applied Materials \& Interfaces}\ }\textbf {\bibinfo {volume}
  {11}},\ \bibinfo {pages} {15489} (\bibinfo {year} {2019})}\BibitemShut
  {NoStop}%
\bibitem [{\citenamefont {Nugroho}\ \emph {et~al.}(2018)\citenamefont
  {Nugroho}, \citenamefont {Darmadi}, \citenamefont {Zhdanov},\ and\
  \citenamefont {Langhammer}}]{NugDarZhd18}%
  \BibitemOpen
  \bibfield  {author} {\bibinfo {author} {\bibfnamefont {F.~A.~A.}\
  \bibnamefont {Nugroho}}, \bibinfo {author} {\bibfnamefont {I.}~\bibnamefont
  {Darmadi}}, \bibinfo {author} {\bibfnamefont {V.~P.}\ \bibnamefont
  {Zhdanov}},\ and\ \bibinfo {author} {\bibfnamefont {C.}~\bibnamefont
  {Langhammer}},\ }\href {https://doi.org/10.1021/acsnano.8b02835} {\bibfield
  {journal} {\bibinfo  {journal} {ACS Nano}\ }\textbf {\bibinfo {volume}
  {12}},\ \bibinfo {pages} {9903} (\bibinfo {year} {2018})}\BibitemShut
  {NoStop}%
\bibitem [{\citenamefont {Ekborg-Tanner}\ \emph {et~al.}(2022)\citenamefont
  {Ekborg-Tanner}, \citenamefont {Rahm}, \citenamefont {Rosendal},
  \citenamefont {Bancerek}, \citenamefont {Rossi}, \citenamefont
  {Antosiewicz},\ and\ \citenamefont {Erhart}}]{EkbRahRos22}%
  \BibitemOpen
  \bibfield  {author} {\bibinfo {author} {\bibfnamefont {P.}~\bibnamefont
  {Ekborg-Tanner}}, \bibinfo {author} {\bibfnamefont {J.~M.}\ \bibnamefont
  {Rahm}}, \bibinfo {author} {\bibfnamefont {V.}~\bibnamefont {Rosendal}},
  \bibinfo {author} {\bibfnamefont {M.}~\bibnamefont {Bancerek}}, \bibinfo
  {author} {\bibfnamefont {T.~P.}\ \bibnamefont {Rossi}}, \bibinfo {author}
  {\bibfnamefont {T.~J.}\ \bibnamefont {Antosiewicz}},\ and\ \bibinfo {author}
  {\bibfnamefont {P.}~\bibnamefont {Erhart}},\ }\href
  {https://doi.org/10.1021/acsanm.2c01189} {\bibfield  {journal} {\bibinfo
  {journal} {ACS Applied Nano Materials}\ }\textbf {\bibinfo {volume} {5}},\
  \bibinfo {pages} {10225} (\bibinfo {year} {2022})}\BibitemShut {NoStop}%
\bibitem [{\citenamefont {Lee}\ \emph {et~al.}(2007)\citenamefont {Lee},
  \citenamefont {Noh}, \citenamefont {Flanagan},\ and\ \citenamefont
  {Luo}}]{LeeNohFla07}%
  \BibitemOpen
  \bibfield  {author} {\bibinfo {author} {\bibfnamefont {S.-M.}\ \bibnamefont
  {Lee}}, \bibinfo {author} {\bibfnamefont {H.}~\bibnamefont {Noh}}, \bibinfo
  {author} {\bibfnamefont {T.~B.}\ \bibnamefont {Flanagan}},\ and\ \bibinfo
  {author} {\bibfnamefont {S.}~\bibnamefont {Luo}},\ }\href
  {https://doi.org/10.1088/0953-8984/19/32/326222} {\bibfield  {journal}
  {\bibinfo  {journal} {Journal of Physics: Condensed Matter}\ }\textbf
  {\bibinfo {volume} {19}},\ \bibinfo {pages} {326222} (\bibinfo {year}
  {2007})}\BibitemShut {NoStop}%
\bibitem [{\citenamefont {Luo}\ \emph {et~al.}(2010)\citenamefont {Luo},
  \citenamefont {Wang},\ and\ \citenamefont {Flanagan}}]{LuoWanFla10}%
  \BibitemOpen
  \bibfield  {author} {\bibinfo {author} {\bibfnamefont {S.}~\bibnamefont
  {Luo}}, \bibinfo {author} {\bibfnamefont {D.}~\bibnamefont {Wang}},\ and\
  \bibinfo {author} {\bibfnamefont {T.~B.}\ \bibnamefont {Flanagan}},\ }\href
  {https://doi.org/10.1021/jp100858r} {\bibfield  {journal} {\bibinfo
  {journal} {The Journal of Physical Chemistry B}\ }\textbf {\bibinfo {volume}
  {114}},\ \bibinfo {pages} {6117} (\bibinfo {year} {2010})}\BibitemShut
  {NoStop}%
\bibitem [{\citenamefont {Mamatkulov}\ and\ \citenamefont
  {Zhdanov}(2020)}]{MamZhd20}%
  \BibitemOpen
  \bibfield  {author} {\bibinfo {author} {\bibfnamefont {M.}~\bibnamefont
  {Mamatkulov}}\ and\ \bibinfo {author} {\bibfnamefont {V.~P.}\ \bibnamefont
  {Zhdanov}},\ }\href {https://doi.org/10.1103/PhysRevE.101.042130} {\bibfield
  {journal} {\bibinfo  {journal} {Phys. Rev. E}\ }\textbf {\bibinfo {volume}
  {101}},\ \bibinfo {pages} {042130} (\bibinfo {year} {2020})}\BibitemShut
  {NoStop}%
\bibitem [{\citenamefont {Rahm}\ \emph {et~al.}(2021)\citenamefont {Rahm},
  \citenamefont {Löfgren}, \citenamefont {Fransson},\ and\ \citenamefont
  {Erhart}}]{RahLofFra21}%
  \BibitemOpen
  \bibfield  {author} {\bibinfo {author} {\bibfnamefont {J.~M.}\ \bibnamefont
  {Rahm}}, \bibinfo {author} {\bibfnamefont {J.}~\bibnamefont {Löfgren}},
  \bibinfo {author} {\bibfnamefont {E.}~\bibnamefont {Fransson}},\ and\
  \bibinfo {author} {\bibfnamefont {P.}~\bibnamefont {Erhart}},\ }\href
  {https://doi.org/10.1016/j.actamat.2021.116893} {\bibfield  {journal}
  {\bibinfo  {journal} {Acta Materialia}\ }\textbf {\bibinfo {volume} {211}},\
  \bibinfo {pages} {116893} (\bibinfo {year} {2021})}\BibitemShut {NoStop}%
\bibitem [{\citenamefont {Nugroho}\ \emph
  {et~al.}(2016{\natexlab{a}})\citenamefont {Nugroho}, \citenamefont {Iandolo},
  \citenamefont {Wagner},\ and\ \citenamefont {Langhammer}}]{Nugroho2016}%
  \BibitemOpen
  \bibfield  {author} {\bibinfo {author} {\bibfnamefont {F.~A.~A.}\
  \bibnamefont {Nugroho}}, \bibinfo {author} {\bibfnamefont {B.}~\bibnamefont
  {Iandolo}}, \bibinfo {author} {\bibfnamefont {J.~B.}\ \bibnamefont
  {Wagner}},\ and\ \bibinfo {author} {\bibfnamefont {C.}~\bibnamefont
  {Langhammer}},\ }\href {https://doi.org/10.1021/acsnano.5b08057} {\bibfield
  {journal} {\bibinfo  {journal} {ACS Nano}\ }\textbf {\bibinfo {volume}
  {10}},\ \bibinfo {pages} {2871} (\bibinfo {year}
  {2016}{\natexlab{a}})}\BibitemShut {NoStop}%
\bibitem [{\citenamefont {Mishra}\ \emph {et~al.}(2022)\citenamefont {Mishra},
  \citenamefont {Passos}, \citenamefont {Marini}, \citenamefont {Xu},
  \citenamefont {Amigo}, \citenamefont {Gowen}, \citenamefont {Jansen},
  \citenamefont {Biancolillo}, \citenamefont {Roger}, \citenamefont
  {Rutledge},\ and\ \citenamefont {Nordon}}]{MISHRA2022116804}%
  \BibitemOpen
  \bibfield  {author} {\bibinfo {author} {\bibfnamefont {P.}~\bibnamefont
  {Mishra}}, \bibinfo {author} {\bibfnamefont {D.}~\bibnamefont {Passos}},
  \bibinfo {author} {\bibfnamefont {F.}~\bibnamefont {Marini}}, \bibinfo
  {author} {\bibfnamefont {J.}~\bibnamefont {Xu}}, \bibinfo {author}
  {\bibfnamefont {J.~M.}\ \bibnamefont {Amigo}}, \bibinfo {author}
  {\bibfnamefont {A.~A.}\ \bibnamefont {Gowen}}, \bibinfo {author}
  {\bibfnamefont {J.~J.}\ \bibnamefont {Jansen}}, \bibinfo {author}
  {\bibfnamefont {A.}~\bibnamefont {Biancolillo}}, \bibinfo {author}
  {\bibfnamefont {J.~M.}\ \bibnamefont {Roger}}, \bibinfo {author}
  {\bibfnamefont {D.~N.}\ \bibnamefont {Rutledge}},\ and\ \bibinfo {author}
  {\bibfnamefont {A.}~\bibnamefont {Nordon}},\ }\href
  {https://doi.org/10.1016/j.trac.2022.116804} {\bibfield  {journal} {\bibinfo
  {journal} {TrAC Trends in Analytical Chemistry}\ }\textbf {\bibinfo {volume}
  {157}},\ \bibinfo {pages} {116804} (\bibinfo {year} {2022})}\BibitemShut
  {NoStop}%
\bibitem [{\citenamefont {Abadi}\ \emph {et~al.}(2015)\citenamefont {Abadi},
  \citenamefont {Agarwal}, \citenamefont {Barham}, \citenamefont {Brevdo},
  \citenamefont {Chen}, \citenamefont {Citro}, \citenamefont {Corrado},
  \citenamefont {Davis}, \citenamefont {Dean}, \citenamefont {Devin},
  \citenamefont {Ghemawat}, \citenamefont {Goodfellow}, \citenamefont {Harp},
  \citenamefont {Irving}, \citenamefont {Isard}, \citenamefont {Jia},
  \citenamefont {Jozefowicz}, \citenamefont {Kaiser}, \citenamefont {Kudlur},
  \citenamefont {Levenberg}, \citenamefont {Man\'{e}}, \citenamefont {Monga},
  \citenamefont {Moore}, \citenamefont {Murray}, \citenamefont {Olah},
  \citenamefont {Schuster}, \citenamefont {Shlens}, \citenamefont {Steiner},
  \citenamefont {Sutskever}, \citenamefont {Talwar}, \citenamefont {Tucker},
  \citenamefont {Vanhoucke}, \citenamefont {Vasudevan}, \citenamefont
  {Vi\'{e}gas}, \citenamefont {Vinyals}, \citenamefont {Warden}, \citenamefont
  {Wattenberg}, \citenamefont {Wicke}, \citenamefont {Yu},\ and\ \citenamefont
  {Zheng}}]{tensorflow2015-whitepaper}%
  \BibitemOpen
  \bibfield  {author} {\bibinfo {author} {\bibfnamefont {M.}~\bibnamefont
  {Abadi}}, \bibinfo {author} {\bibfnamefont {A.}~\bibnamefont {Agarwal}},
  \bibinfo {author} {\bibfnamefont {P.}~\bibnamefont {Barham}}, \bibinfo
  {author} {\bibfnamefont {E.}~\bibnamefont {Brevdo}}, \bibinfo {author}
  {\bibfnamefont {Z.}~\bibnamefont {Chen}}, \bibinfo {author} {\bibfnamefont
  {C.}~\bibnamefont {Citro}}, \bibinfo {author} {\bibfnamefont {G.~S.}\
  \bibnamefont {Corrado}}, \bibinfo {author} {\bibfnamefont {A.}~\bibnamefont
  {Davis}}, \bibinfo {author} {\bibfnamefont {J.}~\bibnamefont {Dean}},
  \bibinfo {author} {\bibfnamefont {M.}~\bibnamefont {Devin}}, \bibinfo
  {author} {\bibfnamefont {S.}~\bibnamefont {Ghemawat}}, \bibinfo {author}
  {\bibfnamefont {I.}~\bibnamefont {Goodfellow}}, \bibinfo {author}
  {\bibfnamefont {A.}~\bibnamefont {Harp}}, \bibinfo {author} {\bibfnamefont
  {G.}~\bibnamefont {Irving}}, \bibinfo {author} {\bibfnamefont
  {M.}~\bibnamefont {Isard}}, \bibinfo {author} {\bibfnamefont
  {Y.}~\bibnamefont {Jia}}, \bibinfo {author} {\bibfnamefont {R.}~\bibnamefont
  {Jozefowicz}}, \bibinfo {author} {\bibfnamefont {L.}~\bibnamefont {Kaiser}},
  \bibinfo {author} {\bibfnamefont {M.}~\bibnamefont {Kudlur}}, \bibinfo
  {author} {\bibfnamefont {J.}~\bibnamefont {Levenberg}}, \bibinfo {author}
  {\bibfnamefont {D.}~\bibnamefont {Man\'{e}}}, \bibinfo {author}
  {\bibfnamefont {R.}~\bibnamefont {Monga}}, \bibinfo {author} {\bibfnamefont
  {S.}~\bibnamefont {Moore}}, \bibinfo {author} {\bibfnamefont
  {D.}~\bibnamefont {Murray}}, \bibinfo {author} {\bibfnamefont
  {C.}~\bibnamefont {Olah}}, \bibinfo {author} {\bibfnamefont {M.}~\bibnamefont
  {Schuster}}, \bibinfo {author} {\bibfnamefont {J.}~\bibnamefont {Shlens}},
  \bibinfo {author} {\bibfnamefont {B.}~\bibnamefont {Steiner}}, \bibinfo
  {author} {\bibfnamefont {I.}~\bibnamefont {Sutskever}}, \bibinfo {author}
  {\bibfnamefont {K.}~\bibnamefont {Talwar}}, \bibinfo {author} {\bibfnamefont
  {P.}~\bibnamefont {Tucker}}, \bibinfo {author} {\bibfnamefont
  {V.}~\bibnamefont {Vanhoucke}}, \bibinfo {author} {\bibfnamefont
  {V.}~\bibnamefont {Vasudevan}}, \bibinfo {author} {\bibfnamefont
  {F.}~\bibnamefont {Vi\'{e}gas}}, \bibinfo {author} {\bibfnamefont
  {O.}~\bibnamefont {Vinyals}}, \bibinfo {author} {\bibfnamefont
  {P.}~\bibnamefont {Warden}}, \bibinfo {author} {\bibfnamefont
  {M.}~\bibnamefont {Wattenberg}}, \bibinfo {author} {\bibfnamefont
  {M.}~\bibnamefont {Wicke}}, \bibinfo {author} {\bibfnamefont
  {Y.}~\bibnamefont {Yu}},\ and\ \bibinfo {author} {\bibfnamefont
  {X.}~\bibnamefont {Zheng}},\ }\href {https://www.tensorflow.org/} {\bibinfo
  {title} {{TensorFlow}: Large-scale machine learning on heterogeneous
  systems}} (\bibinfo {year} {2015}),\ \bibinfo {note} {software available from
  tensorflow.org}\BibitemShut {NoStop}%
\bibitem [{\citenamefont {Loshchilov}\ and\ \citenamefont
  {Hutter}(2019)}]{loshchilov2019decoupled}%
  \BibitemOpen
  \bibfield  {author} {\bibinfo {author} {\bibfnamefont {I.}~\bibnamefont
  {Loshchilov}}\ and\ \bibinfo {author} {\bibfnamefont {F.}~\bibnamefont
  {Hutter}},\ }\href@noop {} {\bibinfo {title} {Decoupled weight decay
  regularization}} (\bibinfo {year} {2019}),\ \Eprint
  {https://arxiv.org/abs/1711.05101} {arXiv:1711.05101 [cs.LG]} \BibitemShut
  {NoStop}%
\bibitem [{\citenamefont {Fredriksson}\ \emph {et~al.}(2007)\citenamefont
  {Fredriksson}, \citenamefont {Alaverdyan}, \citenamefont {Dmitriev},
  \citenamefont {Langhammer}, \citenamefont {Sutherland}, \citenamefont
  {Z{\"{a}}ch},\ and\ \citenamefont {Kasemo}}]{Fredriksson2007}%
  \BibitemOpen
  \bibfield  {author} {\bibinfo {author} {\bibfnamefont {H.}~\bibnamefont
  {Fredriksson}}, \bibinfo {author} {\bibfnamefont {Y.}~\bibnamefont
  {Alaverdyan}}, \bibinfo {author} {\bibfnamefont {A.}~\bibnamefont
  {Dmitriev}}, \bibinfo {author} {\bibfnamefont {C.}~\bibnamefont
  {Langhammer}}, \bibinfo {author} {\bibfnamefont {D.~S.}\ \bibnamefont
  {Sutherland}}, \bibinfo {author} {\bibfnamefont {M.}~\bibnamefont
  {Z{\"{a}}ch}},\ and\ \bibinfo {author} {\bibfnamefont {B.}~\bibnamefont
  {Kasemo}},\ }\href {https://doi.org/10.1002/adma.200700680} {\bibfield
  {journal} {\bibinfo  {journal} {Advanced Materials}\ }\textbf {\bibinfo
  {volume} {19}},\ \bibinfo {pages} {4297} (\bibinfo {year}
  {2007})}\BibitemShut {NoStop}%
\bibitem [{\citenamefont {Nugroho}\ \emph
  {et~al.}(2016{\natexlab{b}})\citenamefont {Nugroho}, \citenamefont {Iandolo},
  \citenamefont {Wagner},\ and\ \citenamefont {Langhammer}}]{NugIanWag16}%
  \BibitemOpen
  \bibfield  {author} {\bibinfo {author} {\bibfnamefont {F.~A.~A.}\
  \bibnamefont {Nugroho}}, \bibinfo {author} {\bibfnamefont {B.}~\bibnamefont
  {Iandolo}}, \bibinfo {author} {\bibfnamefont {J.~B.}\ \bibnamefont
  {Wagner}},\ and\ \bibinfo {author} {\bibfnamefont {C.}~\bibnamefont
  {Langhammer}},\ }\href {https://doi.org/10.1021/acsnano.5b08057} {\bibfield
  {journal} {\bibinfo  {journal} {ACS Nano}\ }\textbf {\bibinfo {volume}
  {10}},\ \bibinfo {pages} {2871} (\bibinfo {year}
  {2016}{\natexlab{b}})}\BibitemShut {NoStop}%
\end{thebibliography}
\end{document}